\documentclass[12pt,A4]{article}
\usepackage[T1]{fontenc}
\usepackage[latin9]{inputenc}
\usepackage{amsmath}
\usepackage{amssymb}
\usepackage{graphicx}
\usepackage{enumitem}
\usepackage[numbers]{natbib}
\usepackage{longtable}

\makeatletter

\@ifundefined{date}{}{\date{}}

\usepackage{setspace}
\usepackage{indentfirst}
\usepackage[T1]{fontenc}
\usepackage{helvet}
\usepackage{ae,aecompl}
\usepackage{sidecap}
\usepackage[section]{placeins}
\usepackage{epsf}

\renewcommand{\small}{\fontsize{11}{13.6pt}\selectfont}

\usepackage{fancyhdr}

\newcommand{\vv}{{\rm v}}
\newcommand{\gb}{{\rm gb}}

\newcommand{\Cr}{{\rm Cr}}
\newcommand{\Co}{{\rm Co}}
\newcommand{\Mn}{{\rm Mn}}
\newcommand{\Fe}{{\rm Fe}}
\newcommand{\Ni}{{\rm Ni}}

\makeatother
\begin{document}

\thispagestyle{empty}

\begin{center}{\bf{\sf \Large A mystery of 'sluggish diffusion' in high-entropy alloys: \\the truth or a myth?}}

\vspace{0.3cm}
{\sf Sergiy V. DIVINSKI$^{1,2,a}$, Alexander POKOEV$^{2,b}$, Neelamegan ESAKKIRAJA$^{3,c}$, and Aloke PAUL$^{3,d}$}

\vspace{0.3cm}
\sf{$^1$Institute of Materials Physics, University of M\"unster, Germany}\\
\sf{$^2$Samara National Research University, Samara, Russia}\\
\sf{$^3$Department of Materials Engineering, Indian Institute of Science, 
Bangalore, India}

\vspace{0.3cm}
\sf{$^a$divin@wwu.de}; \sf{$^b$a.v.pokoev46@mail.ru}; \sf{$^c$esakkirajan@iisc.ac.in}; \sf{$^d$aloke@iisc.ac.in}  
\end{center}

\bigskip{}

\noindent \textbf{Abstract.} High entropy alloys (HEAs) are considered as a novel class of materials with a large number of components (five and more) available in equiatomic or nearly equatomic proportions. One of the characteristic properties 
of HEAs was believed to be so-called 'sluggish' diffusion that should be crucial for intended high-temperature technological applications. The faith on this myth instead of rigorous experimental analysis played such a dominant role that the first set of data on interdiffusion, in fact based on an improper analysis, were cited in hundreds of articles to state the presence of sluggishness of diffusion rates in high entropy alloys. 

In this review, the recent data on atomic diffusion in HEAs are presented and critically discussed. The discussion is focused on tracer diffusion which is already measured dominantly for polycrystalline, but in some cases for single crystalline high-entropy alloys. The radiotracer technique provided a unique access to the diffusion rates of elements in the alloys and in fortunate cases these transport quantities could be measured in a single experiment, as in the case of Co, Cr, Fe and Mn diffusion in the CoCrFeMnNi HEA. 

Alternatively, a rigorous analysis of the interdiffuson experiments, which provide the diffusion rates of chemical species, too, becomes more and more sophisticated for three and more elements in an alloy and it is challenging to derive physically sound quantities from a general multi-component diffusion experiment. Most promising in this case is the diffusion couple technique, especially the so-called pseudo-binary approach. This approach is analyzed with a focus on the applicability and the possible errors induced if up-hill diffusion appears.

In the overview, it is shown that atomic diffusion in HEAs cannot {\it a priori} be considered as sluggish and both atomic interactions as well as correlation effects are responsible for the observed trends. Even if estimated on the same homologous scale, the diffusion retardation induced by a 'high entropy' in FCC crystals is not simply proportional to the number of alloying components and it is shown to be similar to that induced by, e.g., the $L1_2$ ordering in a binary system. Furthermore, the importance of cross-correlations in diffusion of different species in HEAs is highlighted.

\section{Introduction}

\noindent High entropy alloys (HEAs) were introduced as a novel class of multicomponent alloys, which contain constituents in equiatomic or near equiatomic proportions \cite{Yeh, HEA_book}. An increased configuration entropy was first believed to stabilize the solid solution 
structures, though the usage of the HEA concept in development of multi-phase materials was highlighted recently \cite{Liaw}. Moreover, it was recently shown that solely an increased ($\ge 4$, e.g.) number of constituting elements does not 
guarantees the formation of a random solid solution in a multi-principal element alloy and the alloy melting/processing temperatures and the interatomic 
correlations are of high importance \cite{He}. 

While initial research in HEAs 
was driven towards identifying single phase HEAs, it is now realised 
that true single phase HEAs are rare and probably of more 
fundamental interest \cite{rev}. Phase separation (or decomposition) has been 
observed in numerous HEA systems \cite{HEA_book} which could be accompanied by 
ordering \cite{Meshi, Rogal}. The nature and extent of phase decomposition is 
influenced by the temperature and composition of the alloys. As an example, 
AlCoCrFeNi\footnote{Since there is no consensus about the 
way to list the principal elements in HEAs (alphabetic order, periodic table order, atomic percentage, etc), we will use the simple alphabetic order, as e.g. 
CoCrFeMnNi for the Cantor system.} shows a single phase B2 structure at higher 
temperatures, separates 
into a mixture of the B2 and BCC phases as the temperature is decreased or Al 
content is increased \cite{Singh}. The Cantor alloy, CoCrFeMnNi, which has 
a highly stable single phase structure above $900^\circ$C, undergoes 
decomposition on prolonged annealing at lower temperatures which 
could be triggered by mechanical pre-deformation \cite{Pippan, Otto}. 

While the phase separation in HEAs seems inevitable, it is not necessarily 
undesirable. For example, AlCoCrFeNi with a basket-weave structure of B2+BCC 
phases reveals an excellent strength and oxidation resistance \cite{Zhang} or 
dual phase structure in non-equiatomic alloys of the Co--Cr--Fe--Mn--Ni system provided excellent 
strength-ductility combinations \cite{Li}. It is evident that the phase 
separation is determined by component's diffusion rates and in its turn such a 
phase separation influences significantly the diffusion kinetics in HEAs. The 
diffusion rates chiefly determine the deformation behavior at elevated 
temperatures, phase stability, oxidation properties etc., which are key to 
foster exploration of HEAs for high temperature applications. 

Initially, four so-called 'core' effects of the HEAs were postulated 
\cite{HEA_book}, mainly to attract an attention to the field and to mark some 
basic ideas:

\begin{itemize}
 \item {\bf high configurational entropy} which would to promote single phase (simple solid solution) formation;
 \item {\bf 'severe' lattice distortion} due to different atomic sizes of the 
elements which would induce strong lattice strains;
 \item {\bf 'cocktail' effect} which has no strict definition, corresponding to 
a synergy of positive effects from a mixture of different elements with 
different properties;
 \item {\bf 'sluggish' diffusion} due to atomic level variation of 
the individual jump barriers induced by the mixture of different elements leading to decrease in the diffusion rates and stabilize the HEAs from phase 
decomposition or coarsening of nano-precipitates. 
\end{itemize}

These 'core effects' attracted a lot of attention and were examined in detail. 
A critical review of these basic ideas and of the present understanding of 
these features was recently published by Miracle \cite{cors}. 

As it was already mentioned above, a large number of principal elements in an 
alloy does not guarantee the formation of a random solid solution, even at high 
temperatures. The configuration entropy of an alloy does increase with the 
number of main components, but the mixing enthalpy and/or enthalpy 
of formation of intermetallic phases are important, too and they have to be 
carefully analyzed \cite{Senkov}. Moreover, non-equiatomic compositions may 
offer even better properties \cite{Raabe, Li}. Similarly, recent studies, both 
theoretical and experimental ones, do not support the original idea of 'severe' 
lattice distortions \cite{Oh, Mathilde, Owen}. Combined theoretical and experimental 
study \cite{Oh} revealed that the mean strains in CoCrFeMnNi are relatively 
small, below 1\%, but their fluctuations are significant. The hypothesis about the 'cocktail' effect is probably most vaguely defined and difficult to test it rigorously \cite{cors}. However, this hypothesis 'drives' further interest to investigate unexplored compositions.

The present review is focused on the remaining 'core' effect of HEAs in this 
list, i.e. the 'sluggish' diffusion.

Although the 'core' effects were formulated in 2004 \cite{HEA_book}, the first 
study with diffusion couple technique was published in 2013 \cite{Tsai}. The faith on a myth instead of study based on rigorous experimental analysis played such a dominant role that the data produced by Tsai et al. \cite{Tsai} -- in fact based on inproper analysis, as it will unambiguously be shown in this article -- were immediately included in the textbook \cite{HEA_book, HEA-tb} and review articles, see e.g. \cite{Senkov, Liaw2}, which are further cited in hundreds of articles to state the presence of sluggishness of diffusion rates in high entropy alloys. In this review, we will analyze the approach of Tsai et al. \cite{Tsai} extensively, since it played such an important role on creating an impression on diffusion behavior in HEAs.

Table~\ref{tab:listE} presents the currently available experimental data on diffusion in different HEAs.

\begin{footnotesize}
\singlespacing
\begin{longtable}{p{3cm}p{2.5cm}p{10cm}c}
\caption{The available \emph{experimental} studies of diffusion in different high-entropy alloys.}\label{tab:listE}\\
 \hline
 \parbox[t][1.3cm][t]{2.3cm}{System/ \\Composition} & \parbox[t][1.3cm][t]{2.3cm}{Temperature \\or 
temperature \\ interval} & Diffusion technique, determined values and main 
conclusions & Ref. \\
 \hline
 CoCrFeMn$_{0.5}$Ni & 1173 K -- 1373 K & 
\begin{itemize}[noitemsep, topsep=0em, itemindent=-20pt] 
\vspace{-1em}
\item Diffusion couples, quasi-binary approach; 
\item Two interdiffusion coefficients determined from a single composition profile; 
\item Estimations$^*$ of the tracer diffusion coefficients;
\item {\bf Sluggish} diffusion if normalized on $T_m$ 
\item Dominant effect of local activation barriers \end{itemize}  & 
\cite{Tsai} \\
 CoCrFeNi,\;\;\;\;\; CoCrFeMnNi & 1073 K -- 1373 K & 
\begin{itemize}[noitemsep, topsep=0em, itemindent=-20pt] 
\vspace{-1em}
\item Radiotracer method ($^{63}$Ni); 
\item Bulk and grain boundary diffusion contributions in polycrystalline 
materials; 
\item Measurements of true tracer diffusion coefficients;
\item {\bf Non-sluggish} diffusion; 
\item Importance of both local activation barriers and correlation 
effects 
\end{itemize}  & 
\cite{M-Ni} \\
 CoCrFeNi,\;\;\;\;\; CoCrFeMnNi & 673 K -- 1173 K & 
\begin{itemize}[noitemsep, topsep=0em, itemindent=-20pt] 
\vspace{-1em}
\item Radiotracer method ($^{63}$Ni); 
\item Grain boundary diffusion in polycrystalline materials; 
\item Measurements of true tracer diffusion coefficients;
\item {\bf Non-sluggish} grain boundary diffusion
\end{itemize}  & 
\cite{M-GB} \\
 CoCrFeNi,\;\;\;\;\; CoCrFeMnNi & 1073 K -- 1373 K & 
\begin{itemize}[noitemsep, topsep=0em, itemindent=-20pt] 
\vspace{-1em}
\item Radiotracer method ($^{57}$Co, $^{51}$Cr, $^{59}$Fe, $^{54}$Mn); 
\item Bulk and grain boundary diffusion contributions in polycrystalline 
materials; 
\item Measurements of true tracer diffusion coefficients;
\item {\bf Non-sluggish} diffusion; 
\item Importance of both local activation barriers and correlation 
effects 
\end{itemize}  & 
\cite{M-Acta} \\
 CoCrFeNi,\;\;\;\;\; CoCrFeMnNi & 1373 K & 
\begin{itemize}[noitemsep, topsep=0em, itemindent=-20pt] 
\vspace{-1em}
\item Radiotracer method ($^{57}$Co, $^{51}$Cr, $^{59}$Fe, $^{54}$Mn, 
$^{63}$Ni); 
\item Single crystalline materials; 
\item Measurements of true tracer diffusion coefficients;
\item {\bf Non-sluggish} diffusion; 
\end{itemize}  & 
\cite{Daniel} \\
 (CoCrFeMn)$_{1-x}$Ni$_x$ $x=0.92, 0.6, 0.2$ & 1373 K & 
\begin{itemize}[noitemsep, topsep=0em, itemindent=-20pt] 
\vspace{-1em}
\item Radiotracer method ($^{57}$Co, $^{51}$Cr, $^{59}$Fe, $^{54}$Mn, 
$^{63}$Ni); 
\item Bulk and grain boundary diffusion contributions in polycrystalline materials; 
\item Measurements of true tracer diffusion coefficients;
\item {\bf Non-sluggish} diffusion; 
\end{itemize}  & 
\cite{Josua} \\
 CoCrFeNi & 1273 K & 
\begin{itemize}[noitemsep, topsep=0em, itemindent=-20pt] 
\vspace{-1em}
\item Diffusion couples; 
\item Effective interdiffusion coefficients on both sides of the Matano plane;
\item {\bf Sluggish} diffusion$^{**}$
\end{itemize}  & 
\cite{Kulkarni} \\
 AlCoCrFeNi & 1273 K -- 1373 K & 
\begin{itemize}[noitemsep, topsep=0em, itemindent=-20pt] 
\vspace{-1em}
\item Diffusion couples; 
\item Numerical assessment of mobilities; 
\item Estimates of tracer correlation factors 
\item {\bf Sluggish} diffusion \end{itemize}  & 
\cite{Danielewski} \\
 Al$_x$CoCrCuFeNi  $x=1, 1.5, 1.8$ & 1473 K & 
\begin{itemize}[noitemsep, topsep=0em, itemindent=-20pt] 
\vspace{-1em}
\item Radiotracer method ($^{60}$Co); 
\item Diffusion retardation with increased $x$; 
\item {\bf Sluggish} diffusion \end{itemize}  & 
\cite{Nadutov} \\
 CoCrFeNi,\;\;\;\;\; CoFeMnNi,\;\;\;\;\; CoCrFeMnNi & 1350 K & 
\begin{itemize}[noitemsep, topsep=0em, itemindent=-20pt] 
\vspace{-1em}
\item Diffusion couples; 
\item Numerical assessment of mobilities; 
\item {\bf Non-sluggish} diffusion \end{itemize}  & 
\cite{Kusza} \\
 CoCrFeMnNi & 1273 K -- 1373 K & 
\begin{itemize}[noitemsep, topsep=0em, itemindent=-20pt] 
\vspace{-1em}
\item Diffusion couples; 
\item Numerical assessment of mobilities; 
\item {\bf Non-sluggish} diffusion \end{itemize}  & 
\cite{LJ} \\
 CoCrNi,\;\;\;\;\;\;\;\;\;\;\; CoCrFeNi,\;\;\;\;\; CoCrFeNiPd & 1100 K -- 1400 K & 
\begin{itemize}[noitemsep, topsep=0em, itemindent=-20pt] 
\vspace{-1em}
\item Diffusion couples; 
\item Boltzmann-Matano analysis; 
\item {\bf Non-sluggish} diffusion \end{itemize}  & 
\cite{Jin} \\
 \hline
\end{longtable}

\vspace{-9mm}
{\singlespacing
\noindent$^*${\footnotesize There are however issues with the analysis of the interdiffusion data which are discussed in Section \ref{sec:pb}.}

\vspace{-2mm}
\noindent$^{**}${\footnotesize Because of wrong comparison outcome of this manuscript on sluggishness is not comparable\\

\vspace{-10mm}\noindent as discussed in Section \ref{sec:pb}.}
}\end{footnotesize}

\begin{table}[h]
\singlespacing
\caption{The available theoretical assessment of experimental data on diffusion in different high-entropy alloys.}\label{tab:listT}
\begin{center}
\footnotesize
\begin{tabular}{p{2.5cm}p{12cm}c}
 \hline
 \parbox[t][0.5cm][t]{2.3cm}{System} & Main conclusions & Ref. \\
 \hline
 CoCrFeMn$_{0.5}$Ni & 
\begin{itemize}[noitemsep, topsep=0em, itemindent=-20pt] 
\vspace{-1em}
\item Data of Tsai et al. \cite{Tsai} are analyzed; 
\item Statement about {\bf sluggish} diffusion; 
\item Dominant effect of correlation effects \end{itemize}  & 
\cite{Beke} \\
CoCrFeMn$_{0.5}$Ni & 
\begin{itemize}[noitemsep, topsep=0em, itemindent=-20pt] 
\vspace{-1em}
\item Analysis of interdiffusion results for multicomponent alloys by MAA-light$^*$ approach, with application to experimental results presented by Tsai et al. \cite{Tsai}; 
\item Correlation effects are highest for fastest diffusing species (Mn in CoCrFeNiMn) and do not affect the slowest moving element (Ni); 
\item Correlation effects alone are not sufficient to explain sluggish diffusion of all components in HEAs \end{itemize}  & 
\cite{Allnatt} \\
CoCrFeMn$_{0.5}$Ni & 
\begin{itemize}[noitemsep, topsep=0em, itemindent=-20pt] 
\vspace{-1em}
\item Analysis of interdiffusion results presented by Tsai et al. \cite{Tsai} by three random alloy approaches -- Darken, combined Manning and HEa and MAA-light approaches; 
\item Results from random alloy solutions in good agreement with the experimental results; 
\item Very similar self-diffusion coefficients for all components \end{itemize}  & 
\cite{MP} \\
CoCrFeMn$_{0.5}$Ni & 
\begin{itemize}[noitemsep, topsep=0em, itemindent=-20pt] 
\vspace{-1em}
\item Analysis of body-diagonal diffusion couple method in application to HEAs; 
\item Flexibility to avoid 2-phase regions by database validation; 
\item Measurements of HEA diffusivities using the Boltzmann-Matano-Kirkaldy method \end{itemize}  & 
\cite{Morral} \\
 \hline
\end{tabular}
\end{center}

\vspace{-9mm}
{\singlespacing
\noindent$^*${\footnotesize MAA-light denotes a 'light' version \cite{Allnatt} of the Moleko-Allnatt-Allnatt approach \cite{MMA}.}
}
\end{table}

Currently, there exists a large number of both experimental investigations and theoretical assessments of diffusion behavior in HEAs, Tables \ref{tab:listE} and \ref{tab:listT}. As an interesting trend, one may recognize that while the first investigations insisted on sluggishness of diffusion in HEAs, the majority of recent studies provide solid arguments against this premise.  

This review is structured as follows. First, the basic diffusion techniques, the diffusion couple and (radio)tracer ones are presented. The main focus is devoted to the correct procedure of data evaluation, especially for the diffusion couple technique. It is shown that the pseudo-binary concept, as it was introduced by Paul \cite{pseudo-bi} is rigorous and provides physically sound results, if applicable. As a result, single composition-dependent interdiffusion coefficients can be determined from such couples. The compositional deviations from the pseudo-binary concept, e.g. the appearance of up-hill diffusion, indicate non-zero fluxes of the components which are supposed to remain idle and hinder a correct determination of the interdiffusion coefficients. 

The tracer technique is shown to be superior with respect to clear distinguishing of the bulk and short-circuit, e.g. grain boundary, contributions in a single experiment, although providing the diffusion coefficients for a single composition. A recent advances with respect to combination of tracer and chemical diffusion experiments are highlighted. Finally, the unresolved and 'hot' open problems are listed.  

\section{Diffusion couple technique}

\begin{figure}[ht]
\footnotesize 
\begin{minipage}[b]{0.29\textwidth}
a) \includegraphics[height=3cm]{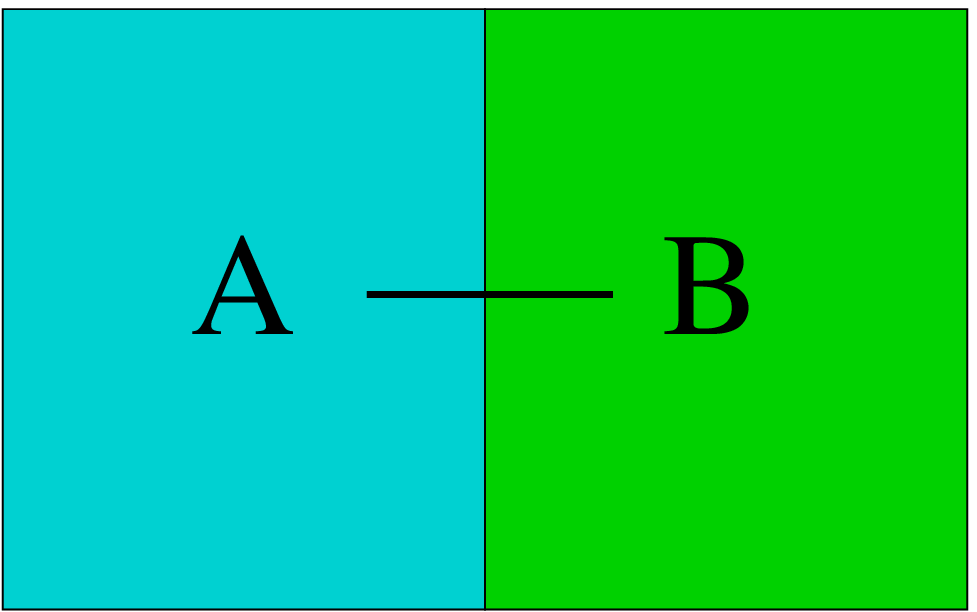}
\end{minipage}
\hfill
\begin{minipage}[b]{0.26\textwidth}
b) \includegraphics[height=3cm]{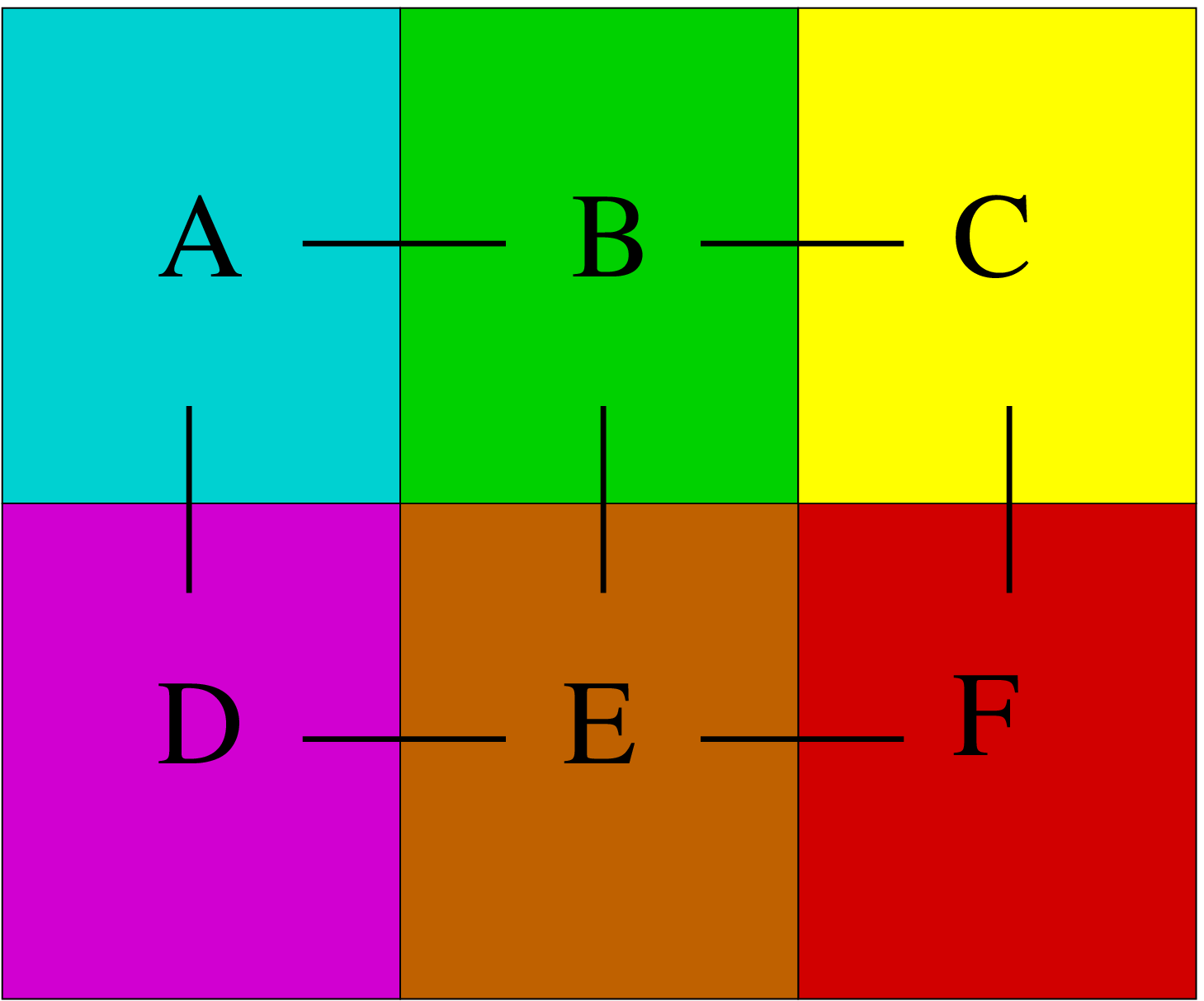}
\end{minipage}
\hfill
\begin{minipage}[b]{0.4\textwidth}
\caption{\footnotesize Schematic arrangements for diffusion couple (a) or diffusion multiples (b) experiments for the investigation of chemical diffusion in multi-component systems. Here A, B, C, D E, F correspond to different alloys. The solid lines indicate the paths of a subsequent chemical analysis.}\label{fig:chemical}\normalsize
\end{minipage}
\end{figure}

\noindent Diffusion in HEAs was dominantly investigated using diffusion couples and/or  multiples \cite{Paul2014}. Probable configurations are schematically shown in Fig. \ref{fig:chemical}.

The first interdiffusion measurements utilizing a diffusion couple technique seemed to confirm the postulated paradigm of sluggish diffusion in HEAs [8,9,11]. 
There was lack of consensus yet on factors responsible for the anticipated decelerated diffusivities. Increased activation barriers [8], correlation effects [9,10], a combination of energy and entropic terms [12] and crystallographic structure [11] have been found to influence the diffusion rates in HEAs. In addition, the diffusion kinetics of only three HEA systems have been analysed so far. 

However, we have to re-analyze the method of processing of the measured concentration profiles, which has to fulfil strinctly the thermodynamic and mathematical restrictions.

\subsection{Conventional method}\label{sec:gen}

\noindent Based on the Onsager formalism, in a system with $n$ components, the interdiffusion flux of $i$th component ($\widetilde{J}_i$) with respect to the interdiffusion coefficients ($\widetilde{D}$) are related by \cite{Paul2014, Daya}

\begin{equation}\label{eq:J-def}
 \widetilde{J}_i = \sum_{j=1}^{n-1} \widetilde{D}^n_{ij} \frac{d C_j}{dx}
\end{equation}

\noindent where $\frac{d C_j}{dx}$ is the concentration gradient of the $j$th component. The interdiffusion fluxes of different components are related by

\begin{equation}
 \sum_{j=1}^{n} \bar{V}_i \widetilde{J}_{i} = 0
\end{equation}

\noindent $\bar{V}_i$ is the partial molar volume of component $i$. $n^{th}$ component is the dependent variable and the interdiffusion flux of this component can be estimated from the known interdiffusion fluxes of other components. The interdiffusion flux of a component $i$ can be estimated from its composition profile utilizing \cite{Paul2014},

\begin{equation}\label{eq:Jt}
 \widetilde{J}(Y^*_{C_i}) = - \frac{C^+_i - C^-_i}{2t} 
 \left[
 \left( 1-Y^*_{C_i} \right) \int^{x^*}_{x^{-\infty}} Y_{C_i} dx +
 Y^*_{C_i} \int_{x^*}^{x^{+\infty}} \left( 1-Y_{C_i} \right) dx +
 \right]
\end{equation}

\noindent where $Y_{C_i} = \frac{C_i - C^-_i}{C^+_i - C^-_i}$ is the concentration normalized variable ($Y^*_{C_i}$ is its value at the given concentration). $C_i$ $\left( = \frac{N_i}{V_m} \right)$ is the concentration of component $i$ ($V_m$ is the molar volume). $C_i^-$ and $C_i^+$ are the concentrations of unaffected left and right hand side of the diffusion couples.

Therefore in a binary system, we have

\begin{subequations}
 \begin{align}
 \widetilde{J}_1 &= - \widetilde{D} \frac{dC_1}{dx} \label{eq:J1}\\
 \widetilde{J}_2 &= - \widetilde{D} \frac{dC_2}{dx} \label{eq:J2}\\
 \bar{V}_1 \widetilde{J}_1 &+ \bar{V}_2 \widetilde{J}_2 = 0 \label{eq:VV}
 \end{align}
\end{subequations}

With the help of standard thermodynamic relation $\bar{V}_1 dC_1 + \bar{V}_2 dC_2 = 0$ \cite{Paul2014}, it can be easily understood from Eqs.(\ref{eq:J1})--(\ref{eq:VV}) that the binary system has only one interdiffusion coefficient at a particular composition estimated at a particular temperature \cite{Paul3}. Combining Equations (\ref{eq:Jt}) and (\ref{eq:J1})--(\ref{eq:VV}), one can estimate the composition dependent interdiffusion coefficients from a single diffusion couple over the whole composition profile. Additionally at the Kirkendall marker plane, one can estimate the intrinsic diffusion coefficients following \cite{Paul4}

\begin{equation}
 D_i = \frac{1}{2t} \left( \frac{\partial x}{\partial C_i} \right)_K
 \left[
 C^+_i \int^{x^K}_{x^{-\infty}} Y_{C_i} dx -
 C^-_i \int_{x^K}^{x^{+\infty}} \left( 1-Y_{C_i} \right) dx
 \right]
\end{equation}

The interdiffusion, intrinsic and tracer diffusion coefficients in a binary system are related by \cite{Paul2014}

\begin{equation}\label{eq:DM}
 \widetilde{D} = C_2 \bar{V}_2 D_1 + C_1 \bar{V}_1 D_2 = 
 \left( N_2 D^*_1 + N_1 D^*_2 \right) \Omega \Phi
\end{equation}

\noindent where $\Phi = \frac{d \ln a_A}{d \ln N_A} = \frac{d \ln a_B}{d \ln N_B}$ is the thermodynamic factor, $a_i$ is the activity of component $i$ and $\Omega$ is the vacancy wind effect \cite{Darken, Manning}. Therefore, we can even estimate the tracer diffusion coefficients if the variation of activity with composition is known.

In a ternary system ($n = 3$), following Equation (\ref{eq:J-def}), the interdiffusion fluxes are related to the interdiffusion coefficients by 

\begin{subequations}
\begin{align}
 \widetilde{J}_1 &= - \widetilde{D}^3_{11} \frac{dC_1}{dx} - 
                     \widetilde{D}^3_{12} \frac{dC_2}{dx} =
                     \widetilde{D}^3_{11} \frac{1}{V_m} \frac{dN_1}{dx} - 
                     \widetilde{D}^3_{12} \frac{1}{V_m} \frac{dN_2}{dx} \label{eq:7a}\\
 \widetilde{J}_2 &= - \widetilde{D}^3_{21} \frac{dC_1}{dx} - 
                     \widetilde{D}^3_{22} \frac{dC_2}{dx} =
                     \widetilde{D}^3_{21} \frac{1}{V_m} \frac{dN_1}{dx} - 
                     \widetilde{D}^3_{22} \frac{1}{V_m} \frac{dN_2}{dx} \label{eq:7b}\\
 \widetilde{J}_1 &+ \widetilde{J}_2 + \widetilde{J}_3 = 0
\end{align}
\end{subequations}

Note that $\widetilde{J}_3$ can be calculated from the estimated values of $\widetilde{J}_1$ and $\widetilde{J}_2$. $\widetilde{D}^3_{11}$ and $\widetilde{D}^3_{22}$ are the main interdiffusion coefficients, and $\widetilde{D}^3_{12}$ and $\widetilde{D}^3_{21}$ are the cross interdiffusion coefficients. Component 3 is the dependent variable \cite{Paul2014, Daya}. Since there are four interdiffusion coefficients to be determined, we need two diffusion couples to intersect at the composition of interest. The intrinsic diffusion coefficients cannot be estimated since the marker plane should be found in both the diffusion couples at the composition of intersection. The diffusion paths follow a serpentine, double serpentine or torturous path depending on many factors and cannot be predicted \emph{a priori} \cite{Paul3}. Therefore, it is almost impossible to design experiments fulfilling such a stringent condition unless it is found incidentally. The second parts of Equations (\ref{eq:7a}) and (\ref{eq:7b}) are true for a constant molar volume, which is considered because of unknown variation of the lattice parameter with composition in a ternary or multicomponent system.

In a system with more than three components, we cannot simply estimate both interdiffusion or intrinsic diffusion coefficients. For example, if we consider a four-component system, the interdiffusion fluxes are related to the interdiffusion coefficients by (keeping 4$^{th}$ component as the dependent variable)

\begin{subequations}
\begin{align}
 \widetilde{J}_i &= - \widetilde{D}^4_{i1} \frac{1}{V_m} \frac{dN_1}{dx} - 
                     \widetilde{D}^4_{i2} \frac{1}{V_m} \frac{dN_2}{dx} -
                     \widetilde{D}^4_{i3} \frac{1}{V_m} \frac{dN_3}{dx}
                     \,\,(i=1,2,3)\label{eq:J-K1} \\
 \widetilde{J}_1 &+ \widetilde{J}_2 + \widetilde{J}_3 + \widetilde{J}_4 = 0 \label{eq:J-K2}
 \end{align}
\end{subequations}

Three interdiffusion fluxes with respect to three dependent component variable ($i =1, 2, 3$) should be written, where component 4 is the dependent variable. Therefore, there are nine interdiffusion coefficients to be determined. This is possible only if three diffusion couples intersect at one particular composition so that total nine interdiffusion fluxes can be related to the interdiffusion coefficients at one particular composition. However,  it is impossible to find the intersection of three diffusion couples at one composition in a four-component space. Therefore, we cannot estimate these parameters in a four-component system. Complications increase even further with the increase in a number of components. 

\subsection{Estimation of the average interdiffusion coefficients}

\noindent As already mentioned, we need two diffusion couples to estimate the interdiffusion coefficients in a ternary system and no data can be estimated in a system with a higher number of components following the conventional method. For the sake of estimation of the data Dayananda and Sohn \cite{D-S} first introduced the concept an average effective diffusion coefficient ($\widetilde{D}_i^{eff}$) from a single diffusion couple. This is expressed for a component $i$ in a $n$-component system as

\begin{equation}\label{eq:Deff}
 \widetilde{J}_i = - \widetilde{D}^{eff}_{i} \frac{dC_i}{dx}
\end{equation}

By comparing Equations (\ref{eq:J-def}) and (\ref{eq:Deff}), we can see that this parameter is related to the main and cross interdiffusion coefficient following a relation 

\begin{equation}\label{eq:Deff2}
 \widetilde{D}^{eff}_i = \sum^{n-1}_{j=1} \widetilde{D}^{n}_{ij} \frac{dC_j}{dC_i}
\end{equation}

This average value of this parameter is estimated by integrating the interdiffusion flux over a composition range of interest such that

\begin{equation}
 \int^{x_2}_{x_1} \widetilde{J}_{i} dx = \int^{C^2_i}_{C^1_i} \widetilde{D}^{eff}_{i} dC_i = \bar{\widetilde{D}}^{eff}_{i} \int^{C^2_i}_{C^1_i}  dC_i =
 - \bar{\widetilde{D}}^{eff}_{i} \left( C^2_i-C^1_i \right)
\end{equation}

\noindent and thus

\begin{equation}\label{eq:Deff3}
 \bar{\widetilde{D}}^{eff}_{i} = - \frac{\int^{x_2}_{x_1} \widetilde{J}_{i} dx}{\left( C^2_i-C^1_i \right)} 
\end{equation}

Since the main and cross interdiffusion coefficients cannot be estimated following this approach, Dayananda and Sohn \cite{D-S} established another modified approach following which one can estimate both of these parameters, although again, these are the average values over a composition range. The main advantage of this approach is that we need only one diffusion couple irrespective of the number of components in a system. In a $n$ component system, after multiplying $(x-x_0)^k$  on both sides of the Matano-Boltzmann relation and then integrating over the composition range of interest \cite{Paul2014, Daya}, we have

\begin{equation}\label{eq:BMav}
 \int^{x_2}_{x_1} \widetilde{J}_{i} (x-x_0)^k dx = \sum^{n-1}_{j=1} \bar{\widetilde{D}}^{n}_{ij} \int^{C_j(x_2)}_{C_j(x_1)} (x-x_0)^k  dC_j
\end{equation}

\noindent where the exponent $k = 0, 1, 2,..,(n-2)$ in a $n$-component system. As explained in Ref. \cite{Paul2014}, in a ternary system, two equations can be written for one component considering $k = 0$ and $1$. Similarly, another two equations can be written for the other component. The third component is considered as the dependent variable. Therefore, four equations can be written for two components to determine the four average interdiffusion coefficients (two main and two cross terms).

Therefore, the average effective diffusion coefficient is an average of the main and the cross interdiffusion coefficients over a composition range in a particular diffusion couple. Some researchers separate a diffusion couple into two parts divided by the Matano plane and then these average values are estimated in these parts for each component. Many others consider the whole composition range for estimating an average. Therefore, these values depend on the random choice of considering the composition range even in a particular diffusion couple.

\begin{figure}[ht]
\begin{center}
\includegraphics[width=0.99\textwidth]{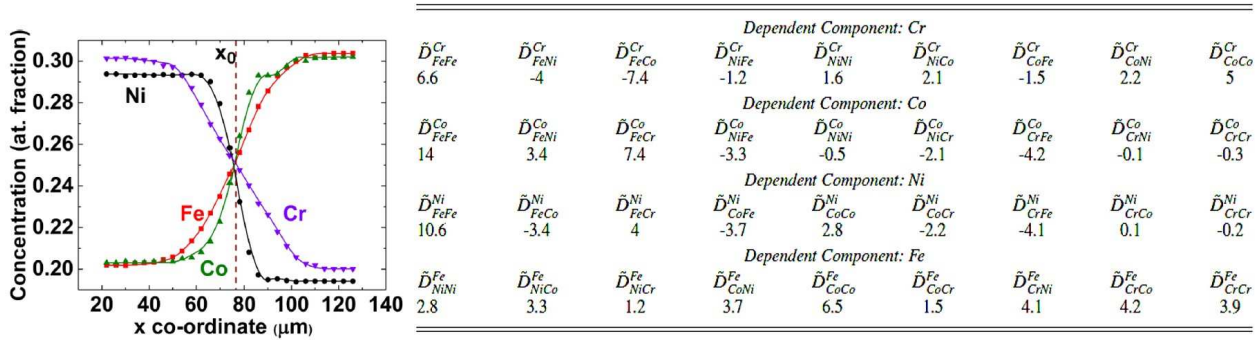}
\end{center}

\caption{\footnotesize The composition profile of Fe$_{0.2}$Co$_{0.2}$Ni$_{0.3}$Cr$_{0.3}$ and Fe$_{0.3}$Co$_{0.3}$Ni$_{0.2}$Cr$_{0.2}$ diffusion couple annealed at 1273 K for 100 hrs and estimated average interdiffusion coefficients ($\times 10^{-16}$ m$^2$/s) \cite{Kulkarni}.}\label{fig:Aloke1}
\end{figure}

Very recently, Kulkarni and Chauhan \cite{Kulkarni} estimated these parameters in a quaternary Fe-Ni-Co-Cr system, see Table\ref{tab:listE}. Note that this is one of the four component base alloy system for two extensively studied five component high entropy alloy systems, where the fifth component is Mn or Al. A diffusion couple was prepared by coupling two alloys with nominal compositions of Fe$_{0.2}$Co$_{0.2}$Ni$_{0.3}$Cr$_{0.3}$ and Fe$_{0.3}$Co$_{0.3}$Ni$_{0.2}$Cr$_{0.2}$ and then annealed at 1273 K for 100 hrs. Therefore, following Equations (\ref{eq:J-K1}) and (\ref{eq:J-K2}), nine interdiffusion coefficients are to be estimated in this quaternary system. Considering $k = 0, 1$ and $2$ for three component (considering the fourth component as the dependent variable), nine equations in total were written for the estimation of the nine average interdiffusion coefficients following Equation (\ref{eq:Deff3}). They followed the similar steps to estimate these set of data considering different components as the dependent variable. The composition profile of the diffusion couple and the data are shown in Fig.~\ref{fig:Aloke1}. These are the average values of the composition range located between $60-104$ $\mu$m, in which the major variation of the composition profiles, are found \cite{Kulkarni}. They reported a possibility of around 40\% error in the calculation.

Kulkarni and Chauhan \cite{Kulkarni} compared the diffusion coefficients estimated in the Fe-Ni-Co-Cr system with the data estimated by Duh and Dayananda in the Fe-Ni-Cr system \cite{Duh}. They stated that since the average diffusion coefficients estimated in the Fe-Ni-Co-Cr system is one order of magnitude lower than the data estimated in Fe-Ni-Cr system (although following the conventional method) [9], it indicates the sluggishness of diffusion coefficient because of alloying addition in a high entropy alloy. However, this comparison was not fully correct. They unfortunately overlooked that the diffusion coefficients in Fe-Ni-Co-Cr are measured at $1000^\circ$C, whereas, the diffusion coefficients in the Fe-Ni-Cr system is measured at $1100^\circ$C. One can easily find an order of magnitude difference in a diffusion coefficient (if not more) with a difference of $100$ K.  Even if the experiments are conducted at the same temperature, one has to be careful before comparison to make sure that the data are measured in a comparable composition range. In different composition ranges, the difference in data by one order of magnitude can easily be found \cite{Paul5}. Therefore, this study wrongly states that diffusion rate decreases because of entropy effect with the increase in a number of components. 

Irrespective of the issues with the comparison of the data in the above-mentioned quaternary system, one may realize the importance of the method for estimation of the average interdiffusion coefficients following the concept proposed by Dayananda and Sohn \cite{D-S}. 

The major advantages are: 

\begin{itemize}
 \item It needs just one diffusion couple to determine both the main and cross interdiffusion coefficients in a system with three or higher number of components.
 \item Therefore, the uncertainties and level of error in estimation of the data could be low even in a ternary system since two diffusion couples are required following the conventional method \cite{Paul2014, Daya}.
 \item One can estimate the diffusion coefficients in four and more number of components, which is simply not possible following the conventional method as discussed in the previous section. 
\end{itemize}

However, one should be aware of the few serious disadvantages of this method:

\begin{itemize}
 \item The estimated average interdiffusion coefficients are not material constants but depend on composition range of a particular diffusion couple. For example, a composition range of 10 at.\% is considered in the Fe-Ni-Co-Cr system \cite{Kulkarni}. Different values will be estimated if different ranges are considered (for example 5, 15 or 20 at.\%) since diffusion paths of this diffusion couple will cover different composition ranges with their own but unknown values over which the average diffusion coefficients are estimated.
 \item The correctness of this approach is always argued based on the back-calculation of the composition profiles of components of a diffusion couple from which these are estimated. However, these data are not useful to calculate back the composition profiles of the components of another diffusion couple prepared with different composition range. 
 \item The values estimated could be confusing. For example, if the diffusion path covers the composition range with both positive and negative cross diffusion coefficients, the average value may be misleading depending on how these influence on the estimated values.
 \item In the example, as shown in Figure \ref{fig:Aloke1}, the diffusion couple was prepared by coupling Fe$_{0.2}$Co$_{0.2}$Ni$_{0.3}$Cr$_{0.3}$ and Fe$_{0.3}$Co$_{0.3}$Ni$_{0.2}$Cr$_{0.2}$. Other diffusion couples can be prepared (even keeping the same composition range of 10 at.\%) by permutations and combinations, for example, Fe$_{0.2}$Ni$_{0.2}$Co$_{0.3}$Cr$_{0.3}$ and Fe$_{0.3}$Ni$_{0.3}$Co$_{0.2}$Cr$_{0.2}$, Fe$_{0.2}$Cr$_{0.2}$Ni$_{0.3}$Co$_{0.3}$ and Fe$_{0.3}$Cr$_{0.3}$Ni$_{0.2}$Co$_{0.2}$. Several other couples can be prepared if we deviate from the body diagonals of a cube but by keeping the same composition range over a sphere around the cube. Even if the diffusion profiles of all the diffusion couples pass through the equiatomic composition, they will produce a completely different set of data because of different composition paths followed by different couples with their unknown diffusion coefficients over which the average values are estimated. Therefore, the estimated data do not represent the actual diffusion coefficients of the equiatomic composition!
 \item Even though it has an advantage that it needs only one diffusion couple, the nature of the Equation (\ref{eq:BMav}) used for estimation of the data introduces unknown error, which increases with the increase in a number of components. A constant molar volume is considered since the variation of the lattice parameters are not known in a multicomponent system. Even if these would be known, it is impossible to locate exactly the position of the Matano plane, $x_0$ \cite{Paul2014}. Therefore, the term $(x-x_0)^k$ introduces an unknown error inherently, which increases drastically with the incrrease in exponent $k$ that is related to the number of components in the system studied. Additionally, in a particular diffusion couple, a set of the equations are written for different $k$ values depending on the number of component. Therefore, different equations will have a different level of errors, which are solved to estimate the data. 
\end{itemize}

\subsection{Pseudo-binary and/or quasi-binary approach}

\noindent The need behind the development of this method was mainly because of the inability of a conventional method to estimate systematic composition dependent diffusion coefficients unless many experiments are conducted in a ternary system. Moreover, there is no possibility of estimation of these data in a system with a higher number of components. The intrinsic diffusion coefficients could not be estimated in a system with more than two components. As explained above, the estimation of the average values by an alternate method does not serve the purpose with meaningful data. 

The concept of the pseudo-binary diffusion couple is established by one of the authors of this manuscript to extend the benefit of a binary system in a multicomponent system \cite{pseudo-bi}. This method produces systematic composition dependent (not average) diffusion coefficients from a single diffusion couple which is otherwise impossible in a ternary or a multicomponent system along with a possibility of estimation of the intrinsic diffusion coefficients. To utilize the benefits of this method, a diffusion couple should be prepared such a way that only two components develop the diffusion profiles keeping other components constant throughout. This removes the complication of the mathematical expressions established based on Onsager formalism, as explained in section \ref{sec:gen}. The benefits of following this method are demonstrated already in few systems \cite{12, 13, 14}. The estimated data even explains the possible change in defects on different sublattices of an intermetallic compound because of alloying, which are otherwise impossible to measure experimentally \cite{13} and explains the effect of alloying in a complicated multicomponent system \cite{15}. 

Immediately after the publication by Paul \cite{pseudo-bi}, Tsai et al. \cite{Tsai} followed this method in CoCrFeMnNi high entropy alloy but named it as the quasi-binary approach. However, their calculations are based on a number of (generally unjustified) approximations, are thus erroneous and worth discussing in detail for the sake of explanation of correct steps one should follow for estimation of the meaningful data. 

They estimated two sets of interdiffusion coefficients for the components which develop the diffusion profiles. Subsequently, Paul \cite{17} wrote an article commenting on this issue based on mathematical and thermodynamical viewpoints that, like in the binary system, a pseudo-binary system also will lead to a single interdiffusion coefficient at a particular composition and temperature. It is crucial to understand that in a strict sense a pseudo-binary system is such, in which two (and only two) main elements develop opposite concentration profiles and there is not flux of any of remaining elements (which concentrations are by construction the same on both ends of the couples). If for any reason there appears a flux of any secondary (hidden) element -- due to cross correlations or inequality of the concentrations at the couple ends -- the system is not a pseudo-binary and one has to be careful with the analysis of the diffusion data. Tsai et al. \cite{Tsai} argued that the deviations from the ideality allows even determination of element-specific interdiffusion coefficients, a fact which is basically wrong for a binary system! Although one may formally introduce element-specific interdiffusion coefficients which would be determined from the corresponding concentration profiles (again, which are identical for a purely binary system!), Paul indicated that these values will have no physical meaning if the pseudo-binary conditions are not fulfilled \cite{17}. 

In a response, Tsai et al. \cite{18} defended their approach saying that not all the pseudo-binary diffusion couples produced diffusion profiles of two components only, but sometimes the components which are kept constant in both the end members of the diffusion couple also produced (uphill) diffusion. In such a situation, a formal application of Equation (\ref{eq:Jt}) to each of two composition profiles will automatically lead to two interdiffusion coefficients one each or every component. In their analysis, the authors of Ref. \cite{MP-HEA} stated that the analysis of Tsai et al. \cite{Tsai} seems to be correct in terms of the random alloy model, if relatively small corrections with respect to the correlation effects are included. We will show below that the analysis of Tsai et al. \cite{Tsai} is basically incorrect and introduces unpredictable uncertainties in the derived values of the interdiffusion coefficients. Thus, one has to be aware of unjustified usage of the corresponding results.  

\begin{figure}[ht]
\begin{center}
\includegraphics[width=0.95\textwidth]{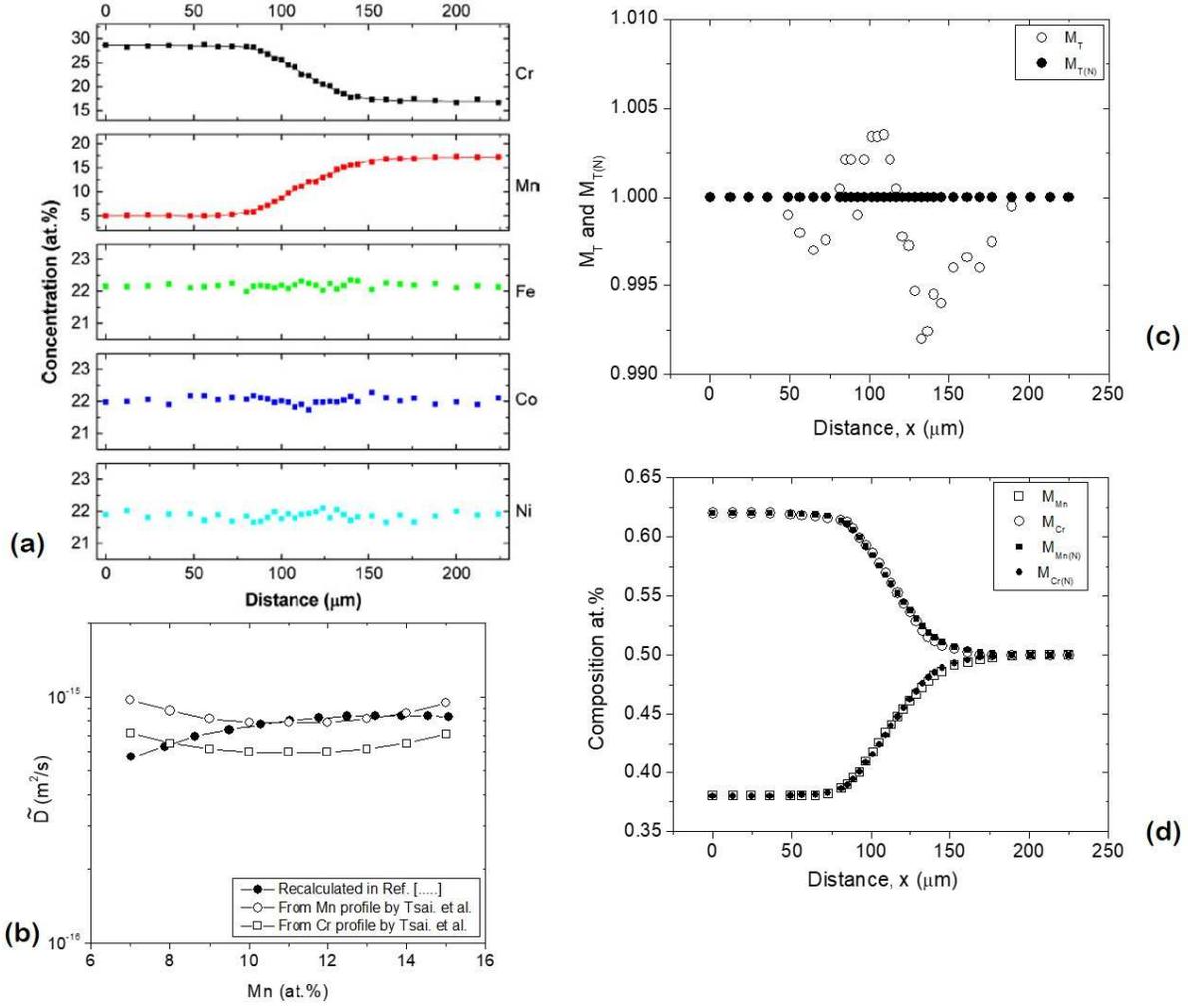}
\end{center}
\caption{\footnotesize (a) Measured composition profiles in (Cr$_{17}$Mn$_{17}$)Co$_{22}$Fe$_{22}$Ni$_{22}$ and (Cr$_{29}$Mn$_{5}$)Co$_{22}$Fe$_{22}$Ni$_{22}$ diffusion couple after annealing at $1000^\circ$C for 100 hrs \cite{Tsai}, (b) the estimated interdiffusion coefficients by Tsai et al. \cite{Tsai} and the corrected calculation by Paul \cite{17} (c) total of modified ($M_T$) and modified normalized composition ($M_T(N)$) profiles (d) modified ($M_i$) and modified normalized composition ($M_i(N)$) profiles of Cr and Mn.}\label{fig:Aloke2}

\end{figure}

\begin{figure}[ht]
\begin{center}
\includegraphics[width=0.95\textwidth]{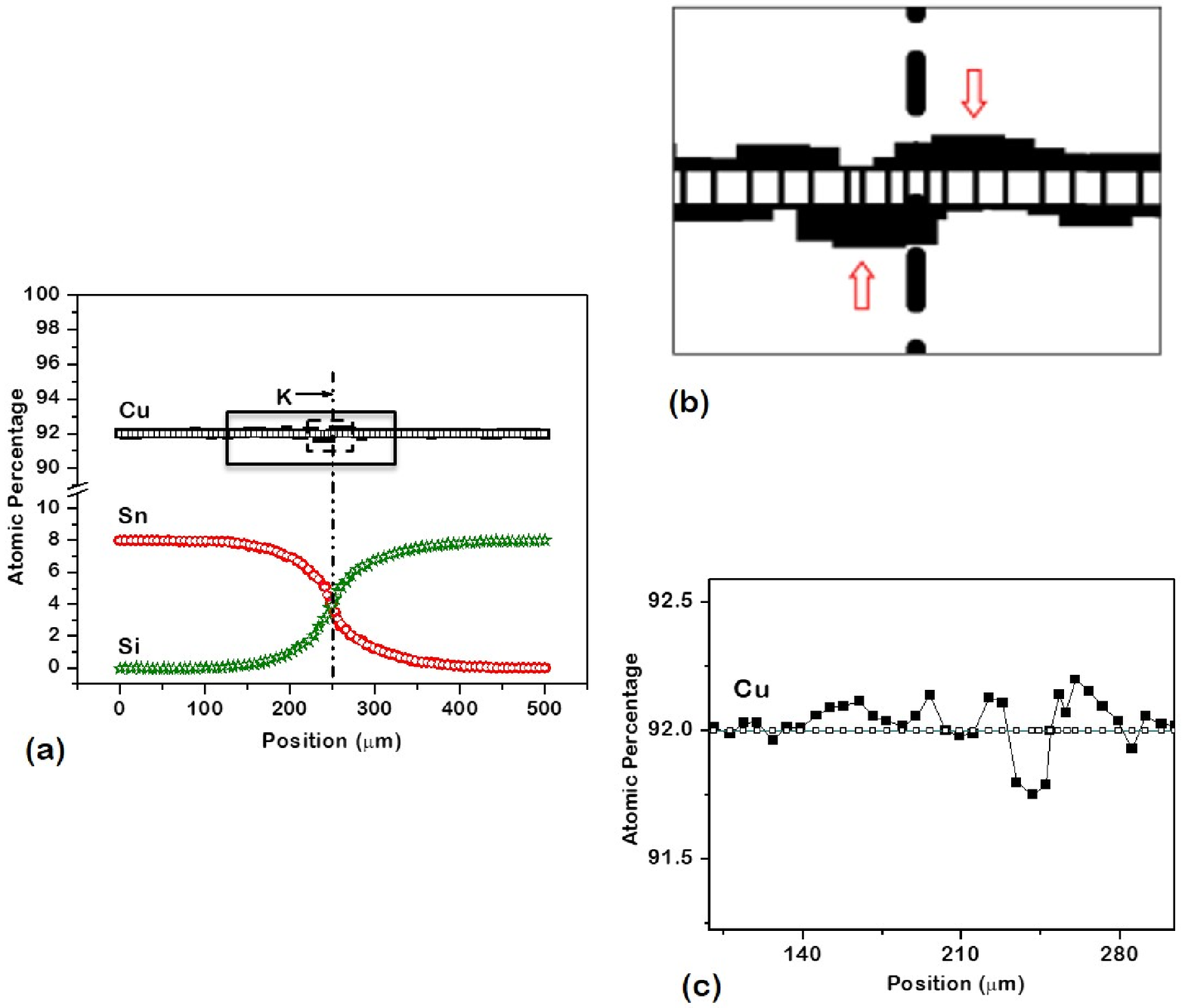}
\end{center}
\caption{\footnotesize (a) Composition profiles of Cu(8at.\%Sn)/Cu(8at.\%Si) diffusion coupled annealed at $700^\circ$C for 25 hrs \cite{pseudo-bi}, (b) a very small part of the Cu profile is enlarged by Tsai et al. \cite{18} to wrongly show this as a proof of an uphill diffusion, (c) an extended part of the same profile is shown as a proof of undulations created because of measurements issues in EPMA. }\label{fig:Aloke3}
\end{figure}

One might mean that the pseudo binary approach developed by Paul \cite{pseudo-bi} and the quasi-binary approach followed by Tsai et al. \cite{Tsai} should be treated as different. In that sense, we could name it as the pseudo-binary approach when ideal diffusion profile is developed and quasi-binary approach when non-ideal diffusion profiles are developed. These lead to several questions, which need to be addressed:

\begin{enumerate}
 \item Tsai et al. \cite{Tsai} have shown the diffusion profile of only one diffusion couple with end member compositions of (Cr$_{17}$Mn$_{17}$)Co$_{22}$Fe$_{22}$Ni$_{22}$ and (Cr$_{29}$Mn$_{5}$)Co$_{22}$Fe$_{22}$Ni$_{22}$ in their original manuscript. This is shown in Figure \ref{fig:Aloke2}a. In this Cr and Mn developed the diffusion profiles and Co, Fe and Ni did not develop any (noticeable) diffusion profiles. The minor undulations at various locations are found because of uncertainties of the composition measurement, which is very common. Therefore, it should produce a single value of the interdiffusion coefficient as it is discussed in detail next.
 \item Tsai with co-workers have not shown the diffusion profiles of other diffusion couples with the uphill diffusion of the components (which are supposed to be constant) in the main article \cite{Tsai}. They mentioned it only in the response \cite{18} to the comments made by Paul \cite{17}. They feel that one should not restrict themselves for the estimation of one interdiffusion coefficient only but estimate two interdiffusion coefficients one for each component. First of all, this behavior is not unusual and will be found frequently in many systems and therefore a detailed discussion on this issue is important.
 \item To strengthen their arguments, Tsai with co-workers cited two other examples in which diffusion coefficients were estimated despite the presence of uphill diffusion of the component (which supposed to be constant) instead of arguing based on mathematical and thermodynamical constraints which should be fulfilled for the estimation of meaningful diffusion coefficients! Otherwise, the data estimated will not have any physical significance and cannot be related to other important basic parameters for understanding the atomic mechanism of diffusion and diffusion-controlled growth mechanism of the interdiffusion zone. Moreover, when a method is followed by someone, it does not justify the correctness of an approach unless validated with scientific logic.
 \item To continue the discussion on this comment, one of such examples cited by Tsai et al. \cite{18} is the experimental results in the Cu(Ga,Sn) solid solution published by Sangeeta and Paul \cite{12} with the intention to show that Paul and coworkers also have followed the pseudo binary approach in such a situation. In support of their comment, they enlarged a very small part (as shown by dotted line rectangle in Figure \ref{fig:Aloke3}a) of the Cu(Sn)-Cu(Ga) diffusion profile to indicate the presence of uphill diffusion Cu, as shown in Figure \ref{fig:Aloke3}b. However, they failed to understand that this is not an uphill diffusion but undulations created by electron probe micro-analysis (EPMA). This must be clear in Figure \ref{fig:Aloke3}c showing a longer part of the diffusion profile (as indicated by a rectangle with a solid line in Figure \ref{fig:Aloke3}a) with many undulations. 
 \item However, the discussion of this particular issue is still valid. As they witnessed the presence of diffusion profile of other components (although not shown in the article by Tsai et al. \cite{Tsai}), what should be the approach of analysis in such a situation? Should it be named differently as the quasi-binary approach which does not develop strictly the diffusion profiles following the concept of the pseudo binary approach proposed by Paul \cite{pseudo-bi, 19}?
\end{enumerate}

To discuss these issues let us first follow the concept of the pseudo binary approach fulfilling the mathematical and thermodynamical restrictions in a multicomponent system. We explain this with the help of diffusion profile (as reported by Tsai et al. \cite{Tsai}) developed in (Cr$_{17}$Mn$_{17}$)Co$_{22}$Fe$_{22}$Ni$_{22}$ and (Cr$_{29}$Mn$_{5}$)Co$_{22}$Fe$_{22}$Ni$_{22}$ diffusion couple, which fulfills the concept such that only Cr and Mn develop the diffusion profiles keeping other components as constant.  Since the other diffusion couples with the uphill diffusion of the other components are not reported by them in the article, we consider different hypothetical diffusion couples for our discussion. For the sake of discussion, these two aspects are discussed in two different sections by naming them as the pseudo binary (as named by Paul \cite{pseudo-bi} and the quasi-binary approach (as named by Tsai et al. \cite{Tsai}). 

\subsection{The pseudo-binary approach (as named by Paul \cite{pseudo-bi, 12, 13, 17})}\label{sec:pb}

\noindent The reason for finding only one and the same interdiffusion coefficient for both the components at a particular composition and temperature in a pseudo-binary diffusion couple should be addressed first. As already explained, based on the relations expressed in Equations (\ref{eq:J1})--(\ref{eq:VV}), one can easily find the reason for the only one interdiffusion coefficient in a binary system. The same logic is true in the pseudo-binary conditions, too. For example, in the (CrMn)CoFeNi five component pseudo-binary diffusion couple considering a constant molar volume (since the variations of lattice parameters in a multicomponent system are not known, but large variations are not expected), the interdiffusion fluxes with respect to the interdiffusion coefficients and the concentration gradients (keeping component $5$ as the dependent variable) can be expressed as

\begin{subequations}
\begin{align}
 \widetilde{J}_{i} &= - \widetilde{D}^{5}_{i1} \frac{1}{V_m} \frac{dN_1}{dx} 
   - \widetilde{D}^{5}_{i2} \frac{1}{V_m} \frac{dN_2}{dx}
   - \widetilde{D}^{5}_{i3} \frac{1}{V_m} \frac{dN_3}{dx}
   - \widetilde{D}^{5}_{i4} \frac{1}{V_m} \frac{dN_4}{dx} (i=1,2,3,4) \\
 & \sum_{j=1}^{5} \widetilde{J}_{i} = 0
 \end{align}
\end{subequations}

Therefore, sixteen interdiffusion coefficients are required to estimate by intersecting four diffusion couples at one composition of interest, which is simply not possible in a 5 component space. Figura \ref{fig:Aloke2}a shows the experimental results of the pseudo-binary diffusion couple in the (CrMn)CoFeNi system, in which only Cr and Mn develop the diffusion profiles keeping Co, Fe and Ni constant throughout the diffusion couple. The small undulations of Co, Fe, and Ni do not indicate a clear presence of the uphill diffusion. In such as situation, we have

\begin{subequations}
 \begin{align}
 \widetilde{J}_{\Cr} &= - \widetilde{D}(\Cr) \frac{1}{V_m} \frac{dN_\Cr}{dx} \\
 \widetilde{J}_{\Mn} &= - \widetilde{D}(\Mn) \frac{1}{V_m} \frac{dN_\Mn}{dx} \\
 &\widetilde{J}_{\Cr} + \widetilde{J}_{\Mn} = 0\label{eq:11c}
 \end{align}
\end{subequations}

\noindent since other components do not develop diffusion profiles, $\widetilde{J}_{\Ni} =\widetilde{J}_{\Co} = \widetilde{J}_{\Fe} = 0$, $\frac{dN_\Co}{dx}=\frac{dN_\Ni}{dx}=\frac{dN_\Fe}{dx}=0$.  Other important relations related to the compositions can be written as 

\begin{equation}
 N_\Mn+N_\Cr+N_\Ni+N_\Co+N_\Fe=1
\end{equation}

At any particular composition in the interdiffusion zone, we can write

\begin{equation}\label{eq:11e}
 dN_\Mn + dN_\Cr = 0
\end{equation}

From the above equations, we have 

\begin{equation}
 \widetilde{D}(\Cr) = \widetilde{D}(\Mn)
\end{equation}

Therefore, we should have the same value at a particular composition irrespective of the composition profile that is considered for the estimation of the interdiffusion coefficient. For estimation of these data, one may normalize compositions as modified compositions \cite{17}

\begin{subequations}
\begin{align}
M_\Cr &=N_\Cr + \frac{1}{2} \left( N_\Ni+N_\Co+N_\Fe \right) \label{eq:12a}\\
M_\Mn &=N_\Mn + \frac{1}{2} \left( N_\Ni+N_\Co+N_\Fe \right) \label{eq:12b}\\
M_\Cr &+M_\Mn=1 \label{eq:12c}
\end{align}
\end{subequations}

Therefore, the interdiffusion coefficient can be directly estimated using the modified composition profiles following (considering constant molar volume \cite{Paul2014})

\begin{equation}\label{eq:13}
 \widetilde{D}(Y^*_i) = \frac{1}{2t} \left( \frac{\partial x}{\partial Y_{M_i}} \right)_{Y^*_i}
 \left[
 \left( 1-Y_{M_i} \right) \int^{x^*}_{x^{-\infty}} Y_{M_i} dx +
 Y_{M_i} \int_{x^*}^{x^{+\infty}} \left( 1-Y_{M_i} \right) dx
 \right]
\end{equation}

\noindent where $Y_{M_i} = \frac{M_i-M_i^-}{M_i^+-M_i^-}$. As explained in Ref. \cite{17}, combining Equations (\ref{eq:11c}), (\ref{eq:11e}) and (\ref{eq:13}), one may easily realize that the same value of the interdiffusion coefficient will be estimated when the composition profile of Cr or Mn is used. Interdiffusion coefficients are a kind of average of the intrinsic diffusion coefficients of the components, which are related by \cite{17}

\begin{equation}\label{eq:14}
 \widetilde{D} = M_\Cr D_\Mn + M_\Mn D_\Cr                                                                                        
\end{equation}

Equation (\ref{eq:14}) also helps to draw an important conclusion. Since the intrinsic diffusion coefficients are estimated at one particular composition, these will have specific values at the annealing temperature. Therefore, the interdiffusion coefficient at a particular composition also will have a single value in a pseudo-binary diffusion couple. Tsai et al. \cite{Tsai} estimated two different values of the interdiffusion coefficients, as shown in Figure \ref{fig:Aloke2}b. In response \cite{18} to the comment \cite{17}, they defended their approach of estimating two different interdiffusion coefficients, one for each component, saying that the (other) couples show the presence of uphill diffusion of components which suppose to remain constant. This is mathematically not correct and induces unpredictable uncertainty in the derived values. However, as discussed in the next section, one can estimate an average interdiffusion coefficient for a composition range of interest following Equation (\ref{eq:Deff}) in such a scenario. Because of thermodynamic and mathematical complications related to the estimation of the diffusion coefficients, one cannot estimate the composition dependent diffusion coefficients from a single diffusion couple. 

To instigate further on this issue, two sentences from the manuscript by Tsai et al. [16] are highlighted:

\emph{There is not much difference in the concentration gradients of Mn and Cr at one particular position. However, the integration term for the estimation of the interdiffusion coefficients (for example in Equation 13) could be high}.

\emph{The total of the interdiffusion fluxes might not be equal to zero leading to two the estimation of two different interdiffusion coefficients when estimated utilizing the composition profiles of different components.} 

These comments indicate that Tsai et al. \cite{Tsai} did not normalize the composition profiles correctly, as explained by Paul \cite{17}. To clarify further, Figure \ref{fig:Aloke2}a shows the composition profiles of the diffusion couple from which the interdiffusion coefficients were estimated. The experimentally measured composition profiles are normalized to the atomic percentage of hundred. However, smooth variations of the profiles which can be directly used for the estimation of the data are not expected because of different level of error at different compositions leading to scattering in data (locally) although with clear evidence of actual variation with respect to the position. As a matter of practice, the diffusion profiles of the components are smoothened separately. After smoothening, one has to make sure that $\sum^n_{i=1} N_i =1$ is fulfilled at every location so that we have $\sum_{i=1}^n \widetilde{J}_i =0$. This was not found by them. The correct steps as explained above only make sure to have the same value of the concentration gradient but with opposite sign. To confirm this doubt, Paul \cite{17} extracted the data from smoothened profiles of Cr and Mn reported by Tsai et al. \cite{Tsai} and then estimated the modified compositions $M_\Cr$  and $M_\Mn$ following Equations (\ref{eq:12a}) and (\ref{eq:12b}). This should ideally lead to the total amount of the two elements, $M_T = M_\Cr + M_\Mn=1$ following Equation (\ref{eq:12c}). As shown in Figure \ref{fig:Aloke2}c, it can be seen that total of the modified compositions $M_T$ at different compositions are indeed not equal to unity. This non-normalized values will automatically give two different values of the interdiffusion coefficients, as indeed found by Tsai et al. \cite{Tsai}, see Figure \ref{fig:Aloke2}b. As a correct step, the modified composition profiles of Cr and Mn should be normalized first by  $M_{\Cr(N)} =\frac{M_\Cr}{M_T}$  and $M_{\Mn(N)} =\frac{M_\Mn}{M_T}$  such that $M_{T(N)}=M_{\Cr(N)}+M_{Mn(N)}=1$, as shown in Figure \ref{fig:Aloke2}c. Following these steps, one can estimate the interdiffusion coefficients using $M_{i(N)}$ instead of $M_i$ in Equation (\ref{eq:13}) to estimate a single value of the interdiffusion coefficient, as shown in Figure \ref{fig:Aloke2}b. Interestingly, this falls just in between the interdiffusion coefficients estimated by Tsai et al. \cite{Tsai} differently for two components. In fact one can directly utilize the normalized composition profiles of $N_{i(N)}$ for the estimation of the interdiffusion coefficient instead of utilizing modified normalized composition profiles $M_{i(N)}$ to get the same values. However, one must utilize $M_{i(N)}$ for the estimation of the intrinsic diffusion coefficients.

\subsection{The quasi-binary approach (as named by Tsai et al. \cite{Tsai})}

Immediately after Paul \cite{pseudo-bi} published the concept behind the pseudo-binary method, Tsai et al. \cite{Tsai} published their experimental work on high entropy alloy and named it as the quasi-binary approach. Tsai et al. \cite{18} argue that we should not restrict ourselves to only two components developing the diffusion profiles strictly in a pseudo (or quasi) binary diffusion couple. This argument is based on the fact that one or more components which are supposed to be kept constant throughout the diffusion couple may also develop (uphill) diffusion profiles in many cases. They argue that in such a situation one can still use this method and estimate different values of composition-dependent interdiffusion coefficients for different components. Therefore, the pseudo-binary approach established by Paul \cite{pseudo-bi, 17} and the quasi-binary approach method followed by them should be treated as different. It is true that such situations are expected to find very frequently in pseudo-binary diffusion couples and therefore a further detailed discussion is important.

\begin{figure}[t]
\begin{minipage}[b]{0.75\textwidth}
 \includegraphics[width=0.99\textwidth]{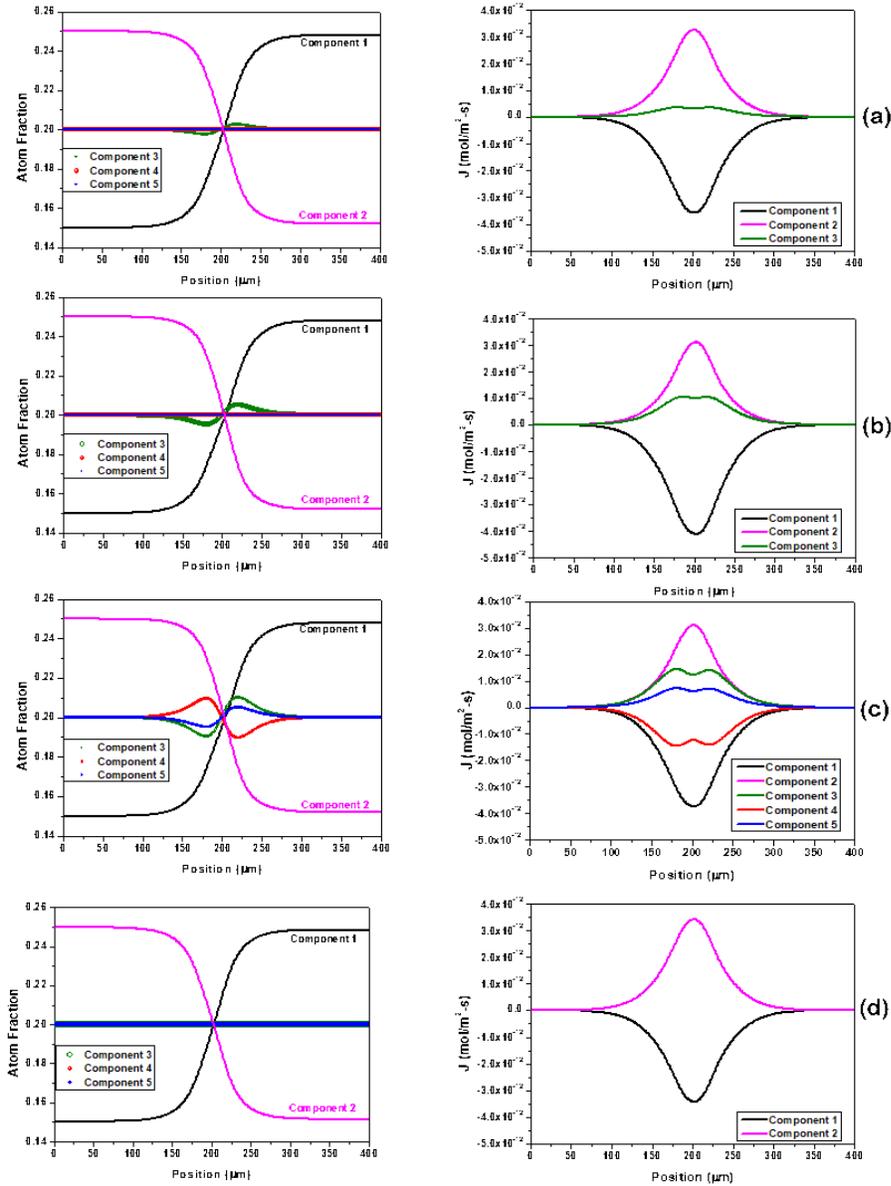}
\end{minipage}
\hfill
\begin{minipage}[b]{0.23\textwidth}
\caption{\footnotesize Non ideal and ideal hypothetical pseudo-binary diffusion profiles are shown. (a) and (b) the uphill diffusion of different extent of component 3, (c) the uphill diffusion of components 3, 4 and 5, (d) ideal variation of the diffusion profiles following the concept of the pseudo-binary diffusion couple.}\label{fig:Aloke4}
\end{minipage}
\end{figure}

To facilitate the discussion, let us consider different hypothetical diffusion couples showing uphill diffusion profiles in a five component system as shown in Figure \ref{fig:Aloke4}. Suppose, as it was recently done \cite{Daniel-BM}, see also below, the diffusion couples are prepared such that components 1 and 2 are used with the composition range of 15-25 at.\% in two end members of the diffusion couple with the expectation that only these two components would develop the diffusion profiles. The composition of other three components 3, 4 and 5 are kept constant in both the end members with the expectation that they remain constant throughout the interdiffusion zone of the diffusion couple. Figures \ref{fig:Aloke4}a--d show different types of situations, which one might find in experiments. Figure \ref{fig:Aloke4}a shows a minor uphill of component 3. We have set the maximum deviation as 5\% of the maximum variation of components 1 and 2 in the end member.  In the next example, in Figure \ref{fig:Aloke4}b, we have given the maximum deviation of 10\% of the same component. There also could be a situation when all the three components which suppose to keep constant but develop diffusion profiles, as shown in Figure \ref{fig:Aloke4}c. In this, we have considered the maximum deviation of component 3 and 4 as 20 percent (in opposite direction) and component 5 as 10\%. In the last, Figure \ref{fig:Aloke4}d shows the ideal variation of the composition profiles following the pseudo binary concept in which only component 1 and 2 develop the diffusion profiles. This type of diffusion profile also will be found frequently as it is reported in few systems already \cite{12,13,14}.  In Paul's group, the ideal variation is found in other systems, too, which are yet to be reported. 

The examples of Figure \ref{fig:Aloke4}a and \ref{fig:Aloke4}b, in which component 3 develops diffusion profile along with component 1 and 2, the diffusion profiles reduce to the situation similar to the ternary system since other two components i.e. component 4 and 5 do not develop the diffusion profiles. These lead to $\frac{dN_4}{dx} =0$, $\frac{dN_5}{dx} =0$, $\widetilde{J}_4 = 0$, $\widetilde{J}_5 = 0$. Therefore, the interdiffusion flux and the interdiffusion coefficients are related by the similar set of equations as expressed in Equation 6 for a ternary system. It further means that cross terms are not zero. Therefore, can we estimate the composition dependent diffusion coefficients from a single diffusion couple like we do in the binary system as it is done by Tsai et al.? The answer is no! In such a situation, if only one diffusion couple is produced instead of two difusion couples by intersecting at the composition of interest (fulfiling the condition for a conventional method of a ternary system), one can measure an average diffusion coefficint following Equation (\ref{eq:Deff3}) or an average of  main and cross interdiffusion coefficients following Equation (\ref{eq:BMav}). If it would be possible to measure the diffusion coefficients as argued by Tsai et al. irrespective of the situation that more than 2 components develop the diffusion profile and still we measure a composition dependent diffusion coefficients from a single diffusion couple, then the diffusion community would not have to struggle for last nine decades to deal with the complexities of the relations in multicomponent system established based on Onsager formalism! One cannot define any types of diffusion parameters unless the mathematical and thermodynamical constraints are fulfilled. Otherwise, the estimated diffusion coefficients will be just a data without any physical significance. A similar argument can be extended to the example as shown in Figure \ref{fig:Aloke4}c in which along with component 1 and 2 other three components i.e. component 3, 4 and 5 also develop the diffusion profiles.

One can estimate the composition-dependent interdiffusion coefficients only if an ideal variation of the diffusion profile is developed as shown in Figure \ref{fig:Aloke4}d. The same method can also be followed when uphill diffusion is witnessed only if the composition profiles are redrawn like Figure \ref{fig:Aloke4}d. It further means that in the nonideal case, the composition profiles first should be replotted by ignoring the uphill diffusion and making them constant throughout the diffusion couple. Following, the composition profiles of component 1 and 2 should be re-plotted such that we have $\sum_{i=1}^n N_i =1$ so that $\sum_{i=1}^n \widetilde{J}_i =0$. By following such a method, we of course introduce an error intentionally, which will vary at different locations. The extent of error roughly can be estimated based on a calculation of the average effective diffusion coefficients estimated following Equation (\ref{eq:Deff3}), as listed in Table \ref{tab:pb}. These are estimated considering the annealing time of 48 hrs. It can be seen that the differences in the data for components 1 and 2 are not very high with respect to the error we expect, in general, in interdiffusion studies \cite{Paul3} since the composition profiles of these components do not change significantly by this modification with such an extent of uphill diffusion considered for this discussion. Additionally, the level of error is different at different locations, which is not captured in the average values. This is an average value for the whole composition range. In a sense when the uphill diffusion is not significant, as shown in Figure \ref{fig:Aloke4}a, the estimation of the composition dependent diffusion coefficient after re-plotting it like Figure \ref{fig:Aloke4}d will not lead much error. The level of error will increase with the increase in uphill diffusion, as shown in Figure \ref{fig:Aloke4}b. However, one should avoid the estimation of the composition dependent interdiffusion coefficients in the situation like in Figure \ref{fig:Aloke4}c. Because of the counterbalance of composition profiles of components 3, 4 and 5, there is not much difference in the average effective diffusion coefficient compared to the ideal variation in Figure \ref{fig:Aloke4}d. However, the influence of the cross terms in this case must be much higher compared to Figure \ref{fig:Aloke4}a and \ref{fig:Aloke4}b. In the case of ideal pseudo binary diffusion couple, as plotted in Figure \ref{fig:Aloke4}d, the average effective interdiffusion coefficients of both the components are same. This will also lead to the same values of the composition dependent interdiffusion coefficients when estimated utilizing the composition profile of any of the components 1 and 2. 

\begin{table}[ht]
 \caption{The average effective interdiffusion coefficients estimated for the diffusion profiles as shown in Figure \ref{fig:Aloke4}.}\label{tab:pb}
\centering

\begin{tabular}{lcc}
\hline
 & \parbox[c][1cm][c]{1cm}{$\widetilde{D}_1^{eff}$} & $\widetilde{D}_2^{eff}$ \\
 & $10^{-15}$ m$^2$/s & $10^{-15}$ m$^2$/s \\
 \hline
 & & \\
 \vspace{0.4cm}
 
 \parbox{12cm}{Case 1 (Figure \ref{fig:Aloke4}a): Component 3 produce uphill diffusion profile. Maximum of uphill is 5\% of the composition difference between components 1 and 2.} & $2.7$ & $3.1$\\
 
 \vspace{0.4cm}
 
 \parbox{12cm}{Case 2 (Figure \ref{fig:Aloke4}b): Component 3 produce uphill diffusion profile. Maximum of uphill is 10\% of the composition difference between components 1 and 2.} & $2.5$ &
 $3.7$\\
 
 \vspace{0.4cm}
 \parbox{12cm}{Case 3 (Figure \ref{fig:Aloke4}c): Component 3 (20\%), 4 (20\%) and 5 (10\%) produce uphill diffusion profiles of the composition difference between components 1 and 2.} & $2.5$ & $3.4$\\
 
 \vspace{0.4cm}
 \parbox{12cm}{Case 4 (Figure \ref{fig:Aloke4}d): Ideal pseudo binary. Only components 1 and 2 produce the diffusion profiles.} & $2.9$ & $2.9$\\
\hline
\end{tabular}
\end{table}

Therefore, we can name it as the pseudo-binary diffusion couple in a situation when only two components develop diffusion profiles keeping the other components constant throughout the diffusion couple. The use of the term \emph{pseudo} (originated from the Greek term 'Pseudes' means 'false') indicates that this is actually a multicomponent system and (falsely) behave like a binary system. Paul \cite{pseudo-bi}, preferred to use this term since it is used very commonly in a multicomponent phase diagram when plotted like a binary phase diagram. On the other hand, the word \emph{quasi} (originated from the Latin term 'Quam si' means 'as if') can be used when the diffusion couple does not develop the diffusion profiles of the components strictly following the concept of the pseudo-binary diffusion couple because of the presence of uphill diffusion or any other reason. Paul personally does not support the use of different names but it is up to the community on the choice of different terms. Instead of creating confusion with different names, we could use the term of ideal and nonideal pseudo-binary diffusion couples.  

Irrespective of the choice of the term by different groups, one can estimate the composition-dependent interdiffusion coefficients only if two components develop diffusion profiles keeping other components more or less the constant within the limit of error of experimental measurements. Such conditions are found already in many systems, like $\beta$-Ni(Pt)Al, Cu(Ga,Sn), Cu(Si,Sn), Fe(Mn)Al successfully \cite{12,13,14}. This is also found in other systems in Paul's group which are yet to be reported. Figure \ref{fig:Aloke2}a also supports the presence of such a profile when Cr and Mn are allowed to diffuse keeping Ni, Co, and Fe constant. However, as mentioned by Tsai et al. \cite{18} other combinations in the same material system show the presence of uphill diffusion. This kind of behavior will also be found frequently. In such a situation, one cannot estimate the composition dependent diffusion coefficients because of the influence of the cross terms on interdiffusion fluxes. Additionally, the estimation of different interdiffusion coefficients for two different components is illogical. One can rather estimate the average diffusion coefficients, as explained in this section. This is an average over a composition range and the values depend on the composition range of a particular diffusion couple. Therefore these are not material constants and do not have any physical significance for understanding the atomic mechanism of diffusion. If one likes to estimate the composition dependent diffusion coefficients, then profiles should be modified such that only two components show the diffusion profiles by making others constant, as explained with the help of Figure \ref{fig:Aloke4}. This will introduce a certain level of error, which will depend on the deviation of the components (showing uphill diffusion) from a fixed value. In case of minor uphill, the error will be negligible and one can go ahead with the calculation of the composition dependent diffusion coefficients following this method. However, the profiles should be normalized correctly, as explained in section 2.3.1, such that the estimated data are physically meaningful to relate to the intrinsic diffusion coefficients correctly following the mathematical and thermodynamical restrictions.

\subsection{Pseudo-ternary diffusion couple}

The pseudo binary diffusion couples can estimate the basic diffusion parameters such as the interdiffusion, intrinsic and tracer diffusion coefficients of components, which are important for understanding the underlying atomic mechanism of diffusion. These parameters also help to examine the role of alloying addition and subsequent effect of configurational entropy. On the other hand, in a multicomponent inhomogeneous system, the crosses along with the main interdiffusion coefficients play a significant role in the evolution of the interdiffusion flux. For example, all the binaries, Co-Fe, Fe-Ni, Fe-Pt, Co-Ni, Co-Pt, Ni-Pt produce normal diffusion profiles at high temperatures. However, many diffusion couples show the presence of uphill diffusion in the Ni-Co-Fe and Ni-Co-Pt ternary systems because of the strong influence of the cross terms. Kulkarni and Chauhan \cite{Kulkarni, 19} have shown the important role of the cross terms in a Fe-Ni-Co-Cr four component system. However, as discussed in section 2.1, these values are average over a certain composition range and also depend on the diffusion path that is followed by a particular diffusion couple depending on the end member compositions. Therefore, these are not suitable for a fair comparison to understand the role of the addition of components on different parameters based on the systematic study. 

Very recently, Esakkiraja and Paul \cite{20} have proposed a pseudo-ternary method, which is basically an extension of the pseudo-binary method \cite{pseudo-bi, 12, 13}. Following this, one can estimate the composition dependent (not average) main and cross interdiffusion coefficients which were never possible after the development of equations related to the multicomponent system based on Onsager formalism many decades before \cite{Paul2014}. In this method, diffusion couples are prepared such that only three components develop the diffusion profile keeping other components as the constant in both the diffusion couples which intersects at the composition of the intersection. In this also the presence of uphill diffusion of the components which suppose to remain constant could create problems. However, it will be helpful for systematic studies on the effect of alloying addition, which are being formulated currently in the laboratory of Paul.

To summarize, only two purely experimental interdiffusion studies are available for the estimation of the diffusion coefficients in high entropy alloys. Kulkarni et al. \cite{Kulkarni} missed the temperature difference and wrongly compared with the data in a ternary system to state that diffusion rates in the quaternary system are sluggish. Tsai et al. \cite{Tsai} made series of errors during estimation of the data. They estimated two different interdiffusion coefficients for two components probably because of the issue of not normalizing the data correctly, as explained in the previous section. Equation (\ref{eq:14}) indicates that this is simply cannot be possible. At one particular composition and temperature, the intrinsic diffusion rates of components are fixed. Therefore, there can be only a single value of the interdiffusion coefficient irrespective the composition profile of the component used in a pseudo-binary diffusion couple. However, other (not main) elements may occasionally develop uphill diffusion. As already explained, in such a situation one can either estimate the average interdiffusion coefficients over a composition range or otherwise, the profiles should be modified to fulfill the mathematical conditions of a pseudo binary diffusion couple for the estimation of the composition dependent interdiffusion coefficients. 

The statement of Tsai with co-workers \cite{Tsai} on the presence of uphill diffusion indicates further vagueness of their calculations. Because of minor movement of the Kirkendall marker plane, they considered the intrinsic diffusion rates of the components are more or less the same, although the exact details of up to what extent these couples developed the uphill diffusion profiles are not mentioned. It should be noted here that in the presence of uphill diffusion because of the influence of the cross terms, the relation between the interdiffusion coefficient and the intrinsic diffusion coefficients are very complicated \cite{Paul2014}.  Even the intrinsic diffusion coefficients will have both the main and cross terms. Additionally, it is known that the calculation of the intrinsic diffusion coefficients is very sensitive to many parameters because of the nature of equation \cite{Paul3}. They further considered that the thermodynamic parameters are negligible in the five component system they studied. This also seems to be not correct \cite{Kulkarni, 19} and their consideration of the intrinsic and tracer diffusion coefficients are the same is not valid. 

Therefore, the study conducted by Tsai et al. \cite{Tsai} is not logically sound to confirm the sluggishness of the diffusion rates of the components because of entropy effect. As discussed in another section, tracer diffusion studies indicate that the addition of a particular component and its effect on other important parameters control the diffusion rates of the components. Entropy effect must be there but it is negligible to control the diffusion rates compared to other more influencing factors. The faith on a myth instead of study based on rigorous experimental analysis played such a dominant role that the data produced by Tsai et al. \cite{Tsai} were immediately included in the textbook \cite{HEA_book} and review articles which are further cited in hundreds of articles to state the presence of sluggishness of diffusion rates in high entropy alloys. 

Therefore, rigorous analyses in many systems are required before drawing any valid conclusion. Just the tracer diffusion study is not enough to understand the underlying diffusion mechanism in an inhomogeneous material system. The recent developments of the concept of the pseudo-binary and pseudo-ternary diffusion couples provide an opportunity to study the multicomponent systems based on the estimation of the composition dependent diffusion coefficients, which were never possible earlier. However, the presence of uphill diffusion of the components which are expected to keep constant throughout the diffusion couple is one of the major drawbacks of these approaches. At present this issue is being further examined in the group of Paul. Most importantly these are the only experimental methods available to estimate the meaningful composition dependent diffusion coefficients in multicomponent inhomogeneous material systems.

\section{Inverse solutions of interdiffusion problem}

As it was mentioned in the Introduction, different types of the numerical inverse solution of the interdiffusion problem, i.e. the determination of the composition-dependent interdiffusion coefficients by fitting the calculated profiles to measured concentration profiles, were suggested recently \cite{LJ, LJ2}. The main idea is very attractive providing the determination of the atomic mobilities (or tracer diffusion coefficients) from the known composition profiles and the thermodynamic relations.

In a binary AB system, the interdiffusion coefficient, $\widetilde{D}$, can directly be determined from a single concentration profile \cite{Paul2014} and it is related to the tracer diffusion coefficients of the elements, $D^*_{\rm A}$ and $D^*_{\rm B}$, via the Darken-Manning relation \cite{Darken, Manning}, Equation (\ref{eq:DM}). It is a common wisdom that knowing the interdiffusion coefficient, one of the tracer diffusion coefficients and the corresponding thermodynamics the missing tracer diffusion coefficient can straightforwardly be determined from Eq.~(\ref{eq:DM}). However, even in a binary system, one may meet an abnormal situation when two different combinations of the tracer diffusion coefficients will represent a solution as it was found for a binary AB system within the framework of the random alloy model below the percolation threshold for a fast diffusing species \cite{MP}. A proper analysis of the correlation factors may help to eliminate a non-physical solution in this particular case \cite{MP}. However, this fact suggests that the numerical inverse method could potentially result in physically wrong results even in the simplest binary couples.

Being exceptional in binary, such a situation may become a more common one in ternary systems, if the concentration of the fast diffusing species will be low, smaller than the percolation threshold. To the best of our knowledge the uniqueness of numerical solution of a such inverse problem -- i.e. the determination of the tracer diffusion coefficients from the known composition profiles and the corresponding thermodynamics -- was not proven so far. 

Nevertheless, as it was mentioned, many efforts were put in the development of numerical procedures which ideally have to allow a direct solution of the inverse diffusion problem providing the concentration-dependent tracer diffusion coefficients from the known chemical profiles and the available thermodynamic assessment of the system \cite{LJ}. According to the authors of the approach, the method was successfully applied to several binary, ternary and multi-component systems \cite{LJ}. However, the authors have not provided a strict proof of the existence of a unique and stable solution of the inverse diffusion problem.

\begin{figure}[t]
\begin{center}
\includegraphics[width=0.95\textwidth]{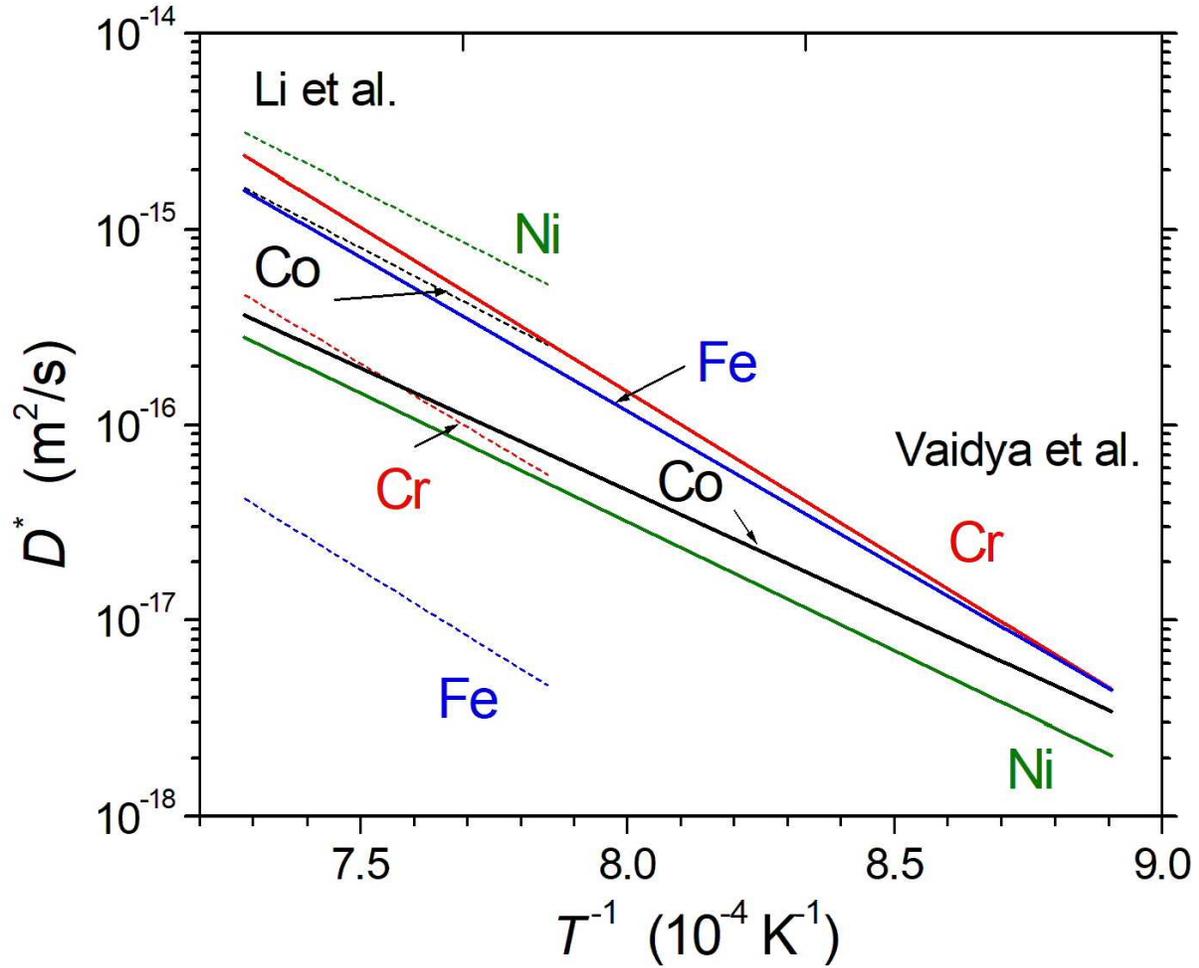}
\end{center}
\caption{\footnotesize Tracer diffusion coefficients directly measured using the radiotracer technique by Vaidya et al. \cite{M-Ni, M-Acta} (solid lines) and determined by the numerical inverse method by Li et al. \cite{LJ} (dashed lines).}\label{fig:Lijun}\normalsize
\end{figure}

The authors \cite{LJ} consider the inverse method as a kind of numerical fitting of the interdiffusion profiles. Note that even a well established tracer method, see below, does not measure the underlying diffusion coefficient of a given chemical element in the compound. Instead, a concentration profile is measured and the tracer distribution is fitted to the experimental data applying (hopefully appropriate) formalism. In a general case it could be involved to disentangle the contributions of volume diffusion and those of various short-circuit paths and time-depending diffusion measurement could be helpful \cite{Paul2014}.

In Fig.~\ref{fig:Lijun} we are comparing the recently published data for the tracer diffusion coefficients in CoCrFeNi determined by the numerical inverse method \cite{LJ}, dashed lines, and the directly measured ones using appropriate radiotracers \cite{M-Ni, M-Acta}, solid lines. The comparison unambiguously substantiate that, at least, one has to use the results of the numerical inverse methods with a care. 
While the following ratio of the tracer diffusion coefficients is established by the direct method, $D^*_{\rm Ni} < D^*_{\rm Co} < D^*_{\rm Fe} < D^*_{\rm Cr}$ \cite{M-Ni, M-Acta}, just opposite trends were found using the numerical inverse method in the work \cite{LJ}, see Fig.~\ref{fig:Lijun}. This example supports the view that one has to use the numerical inverse methods with a care, especially for the multi-component systems, since the solution may be not unique. 

\section{Tracer measurements}

Tracer diffusion technique is a most direct and straightforward method to access the atomic mobilities in homogeneous alloys \cite{Mehrer, Paul2014}. Moreover, several isotopes can be applied simultaneously (with tiny concentrations so that the only driving force is the concentration gradient). 

At the moment, the Cantor alloys, i.e. CoCrFeNi and CoCrFeMnNi, are the most investigated systems with respect to the tracer diffusion \cite{M-Ni, M-GB, M-Acta}. Polycrystalline materials with the grain size of $200~\mu$m or even more were produced by mold-casting and homogenized at $1200^\circ$C for 50~h.

The concentration profiles measured by the parallel mechanical sectioning turned out to be suddenly abnormal for the conditions used, i.e. a coarse-grained material and high diffusion temperatures up to $0.8T_m$ ($T_m$ is the melting point). Since a small amount of the radioactive isotope, $M_0$, was applied to the external surface, one would expect a Gaussian solution of the diffusion problem from an instantaneous source, 

\begin{equation}\label{eq:Gauss}
 \bar{c} = \frac{M_0}{\sqrt{\pi D^* t}} \exp \left( - \frac{x^2}{4D^* t} \right)
\end{equation}

\noindent Here $\bar{c}$ is the tracer concentration which is proportional to the relative specific radioactivity of the section (which is measured in an experiment), $D^*$ the tracer diffusion coefficient and $t$ is the diffusion time. 

\begin{figure}
\small
\begin{center}
\begin{minipage}[b]{0.49\textwidth}
a) \includegraphics[width=0.99\textwidth]{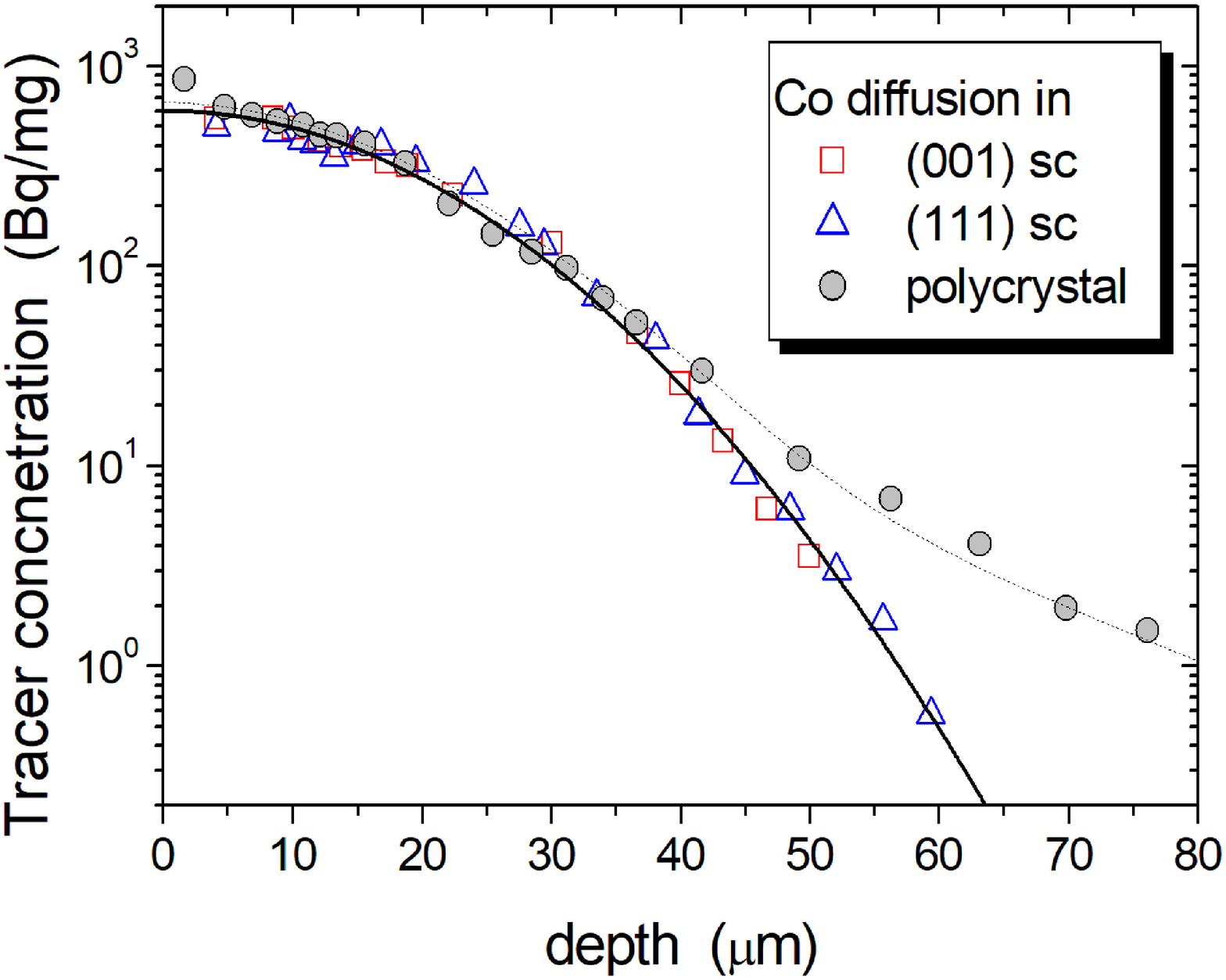}
\end{minipage}
\hfill
\begin{minipage}[b]{0.49\textwidth}
b) \includegraphics[width=0.99\textwidth]{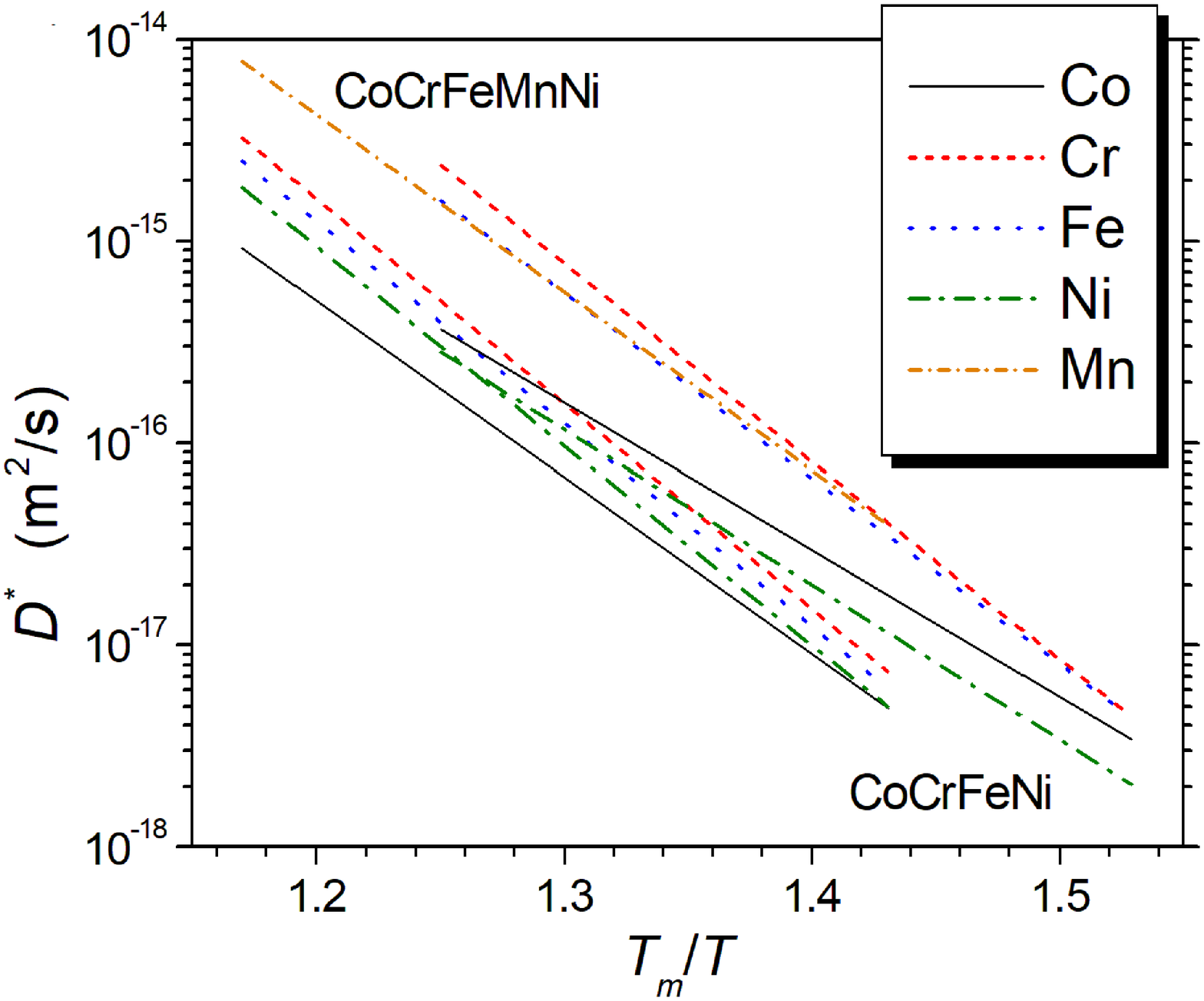}
\end{minipage}
\end{center}
\caption{\footnotesize Examples of the penetration profiles measured for Co diffusion in polycrystalline (circles) and single crystalline (squares and triangles) CoCrFeMnNi alloys (a) and the determined temperature dependencies of tracer diffusion of all constituting elements in polycrystalline CoCrFeNi (measured at lower homologous temperatures) and CoCrFeMnNi (measured at higher homologous temperatures) alloys (b). In (a) Co tracer diffusion was measured along both $\left<001\right>$ (squares) and $\left<111\right>$ (triangles) directions. In (b) the inverse homologous temperature scale, $T_m/T$, is used, with $T_m$ being the melting point of the corresponding compound.}\label{fig:HEAs}\normalsize
\end{figure}

As an example, the measured concentration profile for Co diffusion in CoCrFeMnNi at 1373~K is plotted in Fig.~\ref{fig:HEAs}a, circles. One sees immediately a strong deviation from the expected behavior (shown by solid line) and one may even assume a linear dependence of the tracer concentration against the penetration depth (but not the depth squared, as it has to be according to Eq.~(\ref{eq:Gauss})!)

A most plausible and straightforward explanation (and it turned out to be correct) is that it is the short-circuit diffusion along grain boundaries which contributes to the tracer transport even at such relatively high temperatures and causes this deviation from the Gaussian-like behavior.  Under the particular conditions the grain boundary diffusion proceeds under the so-called B-type kinetic regime after Harrison classification \cite{Har} and one has to expect almost linear dependence in the coordinates of Fig.~\ref{fig:HEAs}a for the corresponding branch of the profile (more precisely a $ln\bar{c} \sim x^{6/5}$ dependence is expected \cite{Paul2014}). A corresponding fit accounting for both the volume and grain boundary diffusion contributions is shown by dashed line in Fig.~\ref{fig:HEAs}a and a good agreement with the measured concentration profile for polycrystalline material is seen.

\begin{table}[ht]
\begin{center}
 \caption{Arrhenius parameters (the pre-exponential factor $D^*_0$ and the activation enthalpy $Q$) of volume diffusion of constituent elements in CoCrFeNi and CoCrFeMnNi HEAs \cite{M-Ni, M-Acta}.}\label{tab:Arr-Cantor}
 
 \begin{tabular}{cccc}
 \hline
 \parbox[c][1cm][c]{1.6cm}{System} & Element & $D^*_0$  (m$^2$/s)  &  $Q$  (kJ/mol)\\
 \hline
 & & & \\
         &  Cr     & $4.2\times 10^{-3}$ & 323 \\
 CoCrFeNi&  Fe     & $4.9\times 10^{-4}$ & 303 \\
         &  Ni     & $1.1\times 10^{-6}$ & 253 \\
         &  Co     & $4.6\times 10^{-7}$ & 240 \\
 & & & \\
 \hline
 & & & \\
         &  Mn     & $1.6\times 10^{-4}$ & 272 \\
         &  Cr     & $2.4\times 10^{-3}$ & 313 \\
 CoCrFeMnNi &  Fe  & $1.3\times 10^{-3}$ & 309 \\
         &  Co     & $1.6\times 10^{-5}$ & 270 \\
         &  Ni     & $6.2\times 10^{-4}$ & 304 \\
 \hline 
 \end{tabular}
\end{center}
\end{table}

However, a direct proof of the correctness of the profile analysis would be the measurements of tracer diffusion in a single crystalline material. Recently, we performed such experiments and measured diffusion of all constituting elements (including solute diffusion of Mn and Cu in CoCrFeNi) in CoCrFeNi and CoCrFeMnNi at 1373~K \cite{Daniel}. The single crystalline HEAs were produced in the research group of Prof. Yu.I. Chumlyakov (Tomsk State University, Russia) by the modified Bridgman technique as massive rods of $24$~mm in diameter. Tracer diffusion of all constituting elements was measured along both $\left<001\right>$ (squares) and $\left<111\right>$ (triangles) directions. In Fig.~\ref{fig:HEAs}a the penetration profiles for Co in CoCrFeMnNi are shown in comparison to the profiles measured on polycrystalline materials for exactly the same diffusion annealing treatment.

Two salient features are immediately seen.

\begin{enumerate}
 \item Diffusion of Co (as well as of all other elements) is isotropic (does not depend on the diffusion direction) as it should be for the crystals of cubic symmetry.
 \item A perfect agreement of the concentration profiles measured in polycrystalline and single crystalline alloys proves the analysis made in Refs.~\cite{M-Ni, M-Acta} with respect to the subdivision of the total penetration profiles in two branches corresponding to bulk and grain boundary contributions.  
\end{enumerate}

Moreover, tracer diffusion in Ref.~\cite{Daniel} was measured for different diffusion times and constant, time-independent diffusion coefficients were found.

\begin{figure}[t]
\begin{center}
\includegraphics[width=0.8\textwidth]{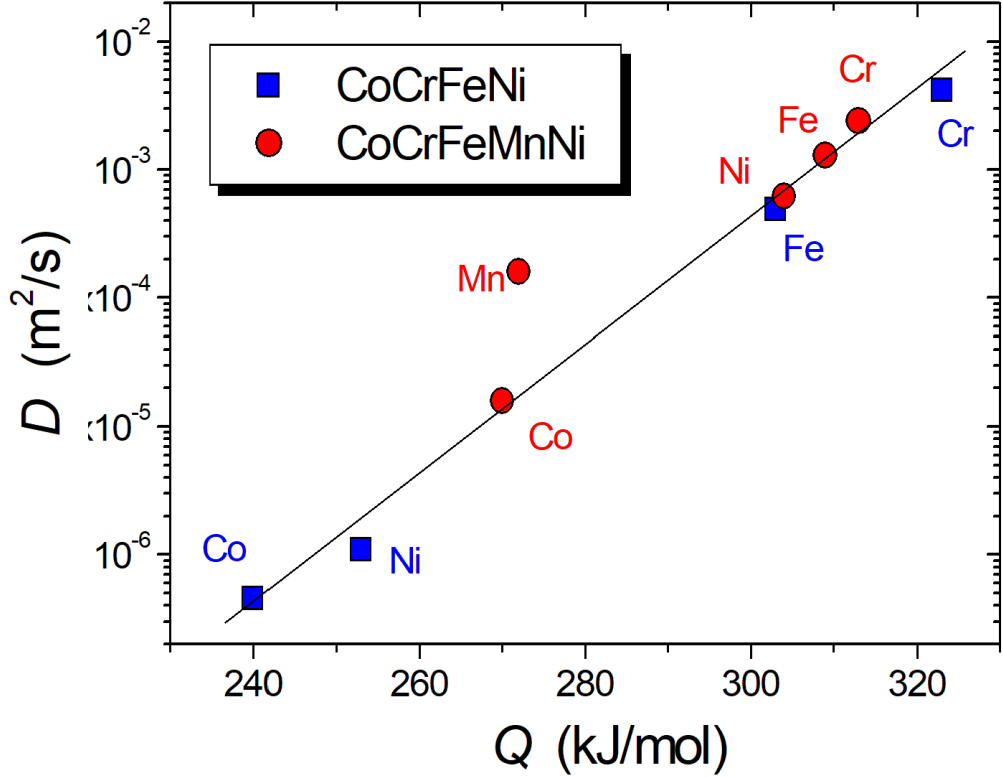}
\end{center}
\caption{\footnotesize Compensation plot for tracer diffusion coefficients of elements in CoCrFeNi and CoCrFeMnNi \cite{M-Ni,M-Acta}.}\label{fig:compensation}\normalsize
\end{figure}

Tracer diffusion in polycrystalline CoCrFeNi and CoCrFeMnNi alloys were measured in the temperature range from 1023 to 1373~K and linear Arrhenius-like behavior was seen \cite{M-Ni, M-Acta}. The determined Arrhenius parameters, i.e. the pre-exponential factors $D^*_0$ and the activation enthalpies $Q$, are listed in Table~\ref{tab:Arr-Cantor}.

In Fig.~\ref{fig:compensation} the so-called compensation plot for diffusion in HEAs is shown, i.e. the pre-exponential factors, $D^*_0$, are plotted against the corresponding activation enthalpies, $Q$. Generally, a nice linear dependence is seen, with one exception for Mn diffusion in CoCrFeMnNi. Nevertheless, it is important to underline that this graph is not a proof that, e.g., $D^*_0$ depends on $Q$. Such a correlation supports an idea that an increase of the activation enthalpy -- the energy barriers for the given atom-vacancy exchange -- is 'compensated' by a corresponding increase of the pre-exponential factor (or corresponding diffusion entropy).

Astonishingly, almost all data points lie on a single line; only Mn demonstrates a strong deviation from the general trend. This behavior may be related to a strong attraction between a vacancy and Mn atoms in the CoCrFeMnNi alloy.

\begin{figure}[ht]
\begin{center}
\includegraphics[width=0.8\textwidth]{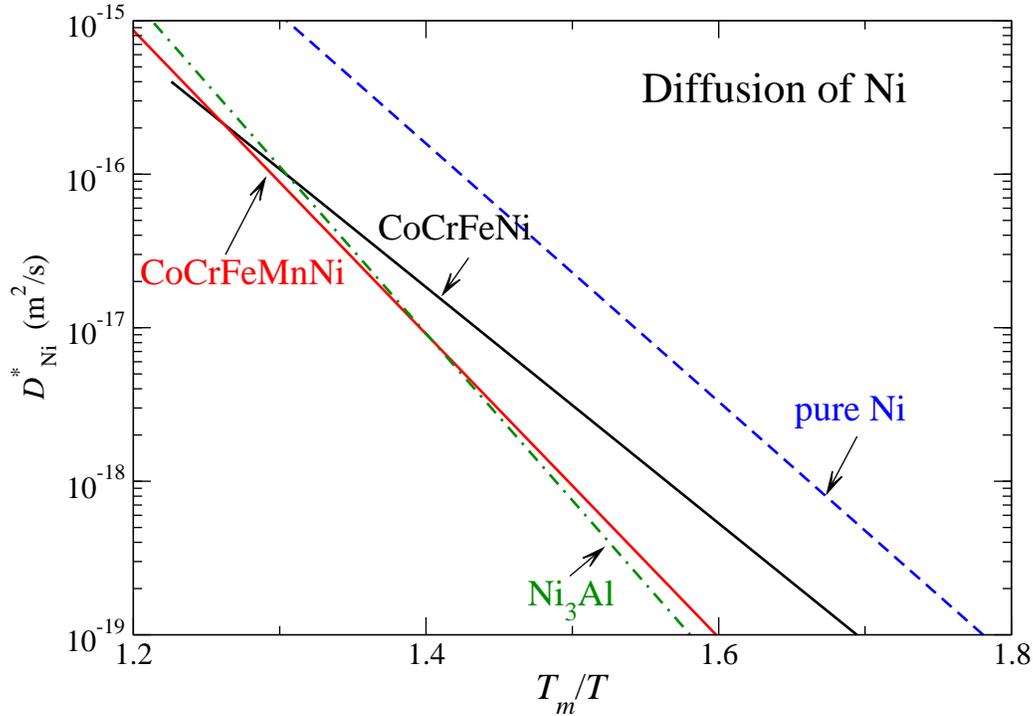}
\end{center}
\caption{\footnotesize Tracer diffusion coefficients, $D^*_{\rm Ni}$, of Ni in CoCrFeNi \cite{M-Ni} (black solid line), CoCrFeMnNi \cite{M-Ni} (red solid line), pure Ni \cite{Ni} (blue dashed line), and stoichiometric $L1_2$-ordered Ni$_3$Al alloy \cite{Ni3Al} (green dot-dashed line) as function of the inverse homologous temperature scale, $T_m/T$. $T_m$ is the melting point of the corresponding compound.}\label{fig:order}\normalsize
\end{figure}

Figure~\ref{fig:HEAs}b suggests a certain retardation of volume diffusion of Ni in FCC systems with equiatomic alloying of subsequent elements. In Fig.~\ref{fig:order} this diffusion retardation is compared to that induced by the $L1_2$ ordering in the Ni$_3$Al alloy. This comparison substantiates that the retardation -- observed when the homologous temperature scale is used -- is not abnormal and is similar to that induced by the sublattice ordering in FCC matrix.  One may speculate that the slow diffusing species, e.g. Co and Ni in CoCrFeMnNi, reveal a tendency to ordering, following the idea that Al diffuses slower than Ni in Ni$_3$Al \cite{Div-ch10}. Alternatively, one may assume that the fast diffusing Mn atoms form a percolating cluster for vacancy transport and win a competition with Ni atoms that retards diffusion of the latter.

\section{Combination of tracer and chemical diffusion}

As it was shown above, a standard tracer diffusion measurement can provide the tracer diffusion coefficients for the given composition of the multicomponent alloy, while the diffusion couple method (according to, e.g., the pseudo-binary approach) yields the interdiffusion coefficient as a function of the composition along the diffusion path. A high-throughput determination of tracer mobilities for a large number of compositions is still illusive for a radiotracer technique and one is limited by a relatively low number of inspected compositions within the corresponding phase diagram, typically by a factor of ten, see e.g. \cite{CuFeNi}.

Recently, a new approach has been suggested which corresponds to a combination of the radiotracer and inter-diffusion measurements and provides the tracer diffusion coefficients of a component as a function of the composition along the diffusion path \cite{Murch1, Murch2}.

\begin{figure}[ht]
 \begin{minipage}[c]{0.35\textwidth}
  \includegraphics[height=4cm]{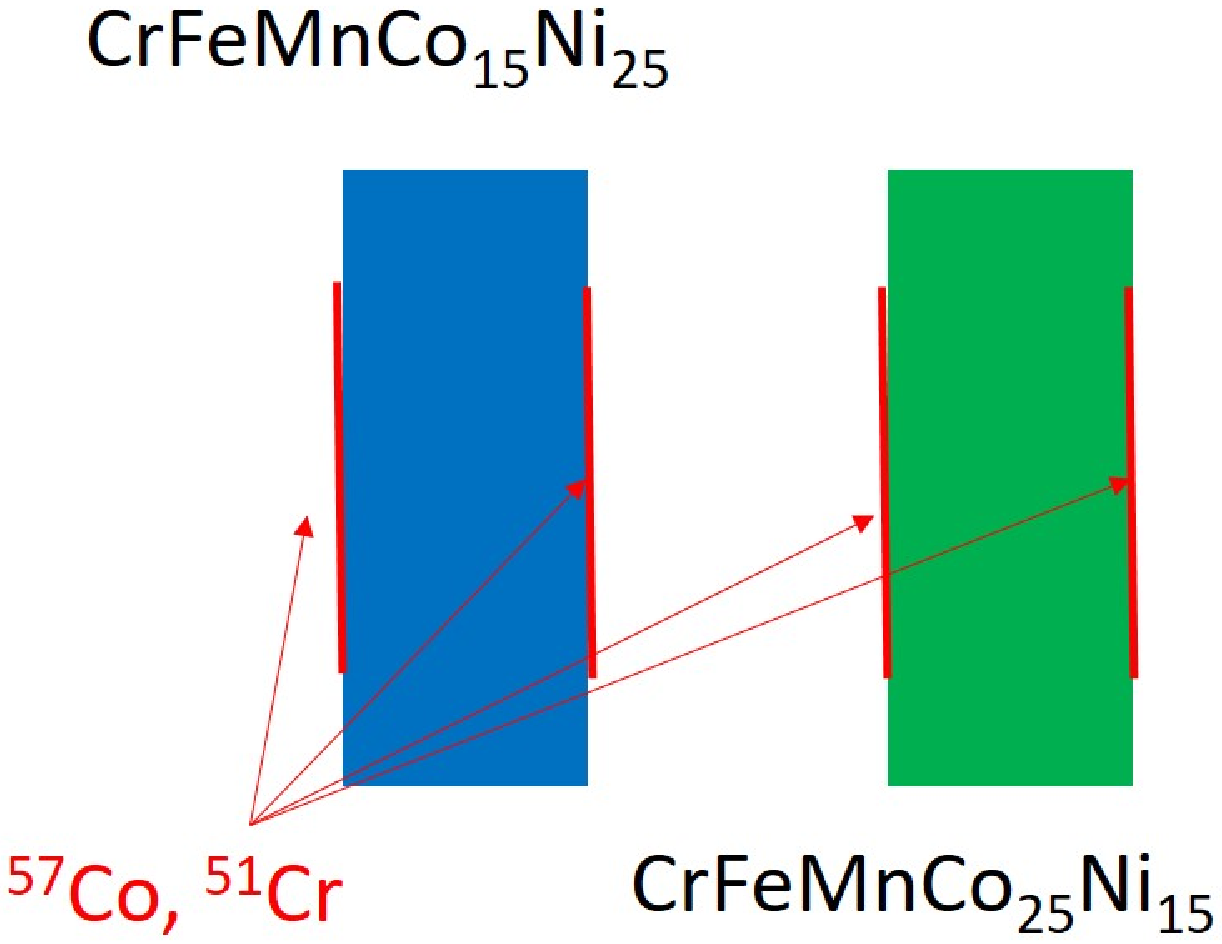}
 \end{minipage}
 \hfill
 \begin{minipage}[c]{0.2\textwidth}
  \includegraphics[height=3cm]{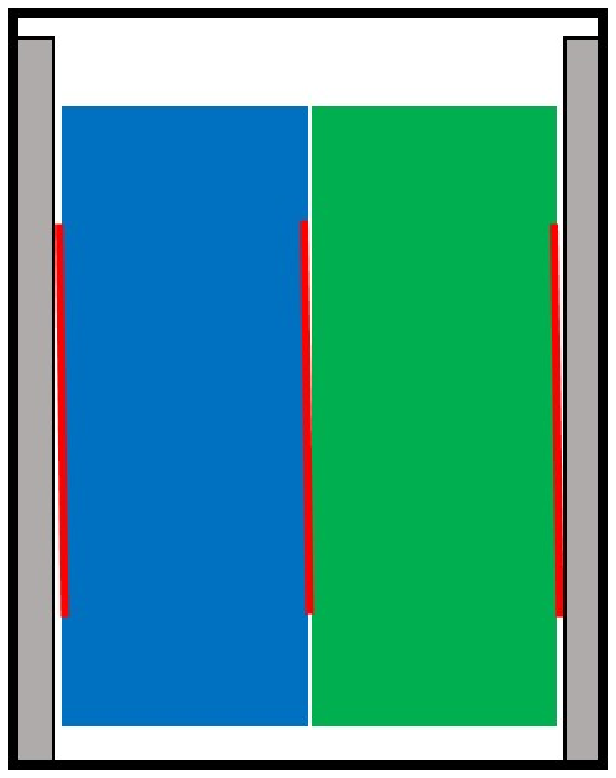}
 \end{minipage}
 \hfill
 \begin{minipage}[c]{0.4\textwidth}
  \includegraphics[height=3cm]{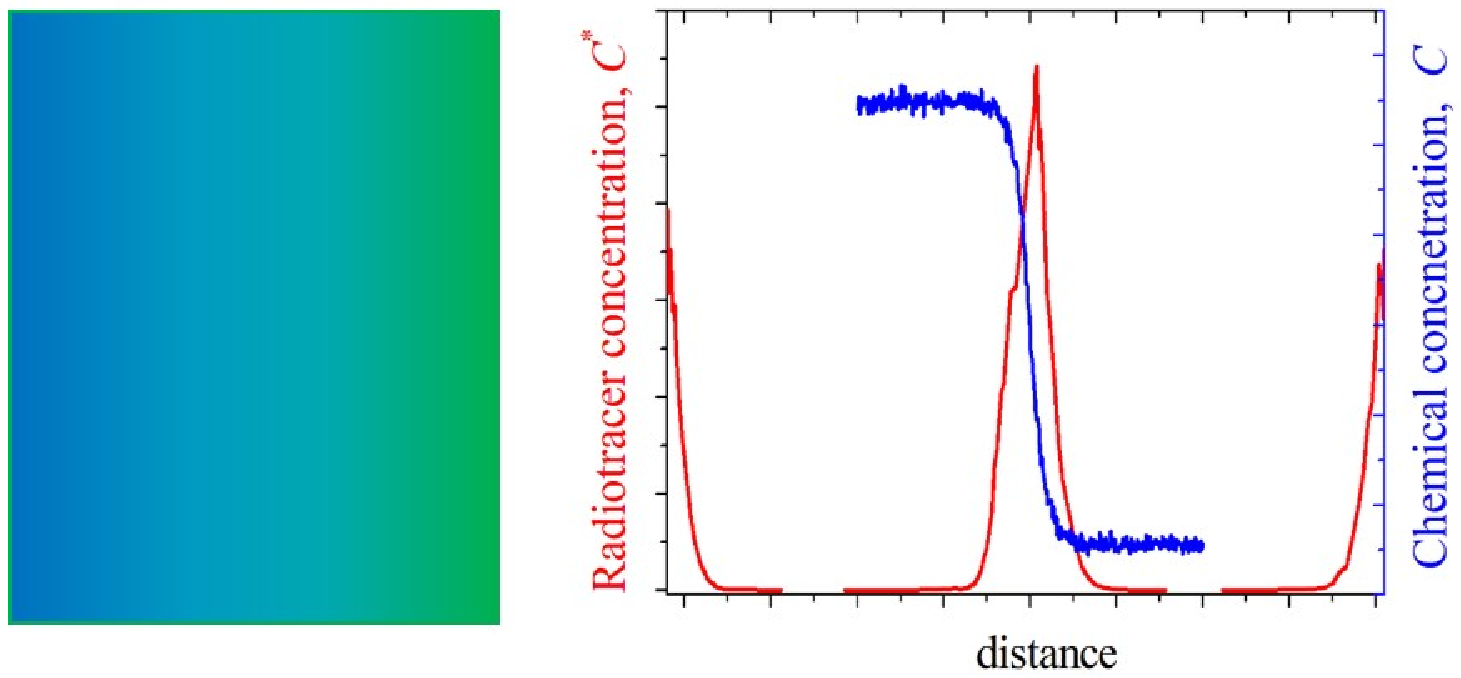}
 \end{minipage}
 
  \caption{\footnotesize A scheme of a modified tracer-interdiffusion couple: two end-members of a pseudo-binary diffusion couple are prepared and a tracer (or tracer's mixture) is deposited on all surfaces (left panel); the couple is mounted in a fixture for interdiffusion studies and subjected to an annealing treatment (center panel); mechanical sectioning combined with chemical profiling -- on two different but identical couples -- results in two profiles, i.e. one for the tracer concentration, $C^*(x)$, red line, and another for the chemical concentration, $C(x)$, blue line (right panel).}\label{fig:comb}
\end{figure}

As a further development, we proposed an extension of this approach suggesting to include the tracer deposition at the both ends of the diffusion couple  \cite{Daniel-BM}. The technique is schematically illustrated in Fig.~\ref{fig:comb} for the case of a pseudo-binary tracer-interdiffusion couple:

\begin{itemize}
 \item Buttons of, e.g., CrFeMnCo$_{15}$Ni$_{25}$ and CrFeMnCo$_{25}$Ni$_{15}$ alloys are plain-parallel ground, polished and a tracer (or tracer solution, e.g. a mixture of $^{51}$Cr, $^{54}$Mn, $^{57}$Co and $^{59}$Fe, which activities could conveniently be discriminated by a suitable $\gamma$-detector) is applied to all surfaces, Fig.~\ref{fig:comb}a;
  \item The samples are tightly placed together in a special fixture and annealed at the given temperature for the given time, Fig.~\ref{fig:comb}b;
  \item In fact, two identical couples are annealed, one with and one without tracer application;
  \item The radioactively contaminated couple is mechanically sectioned and the tracer concentration is measured through the whole couple, the red line in Fig.~\ref{fig:comb}c;
  \item Chemical diffusion in the identical couple is measured by electron probe micro-analysis, the blue line in Fig.~\ref{fig:comb}c.
\end{itemize}

The key moment is the determination of the two concentration profiles, see Fig.~\ref{fig:comb}c, corresponding to the tracer atom concentration, $C^*(x)$, and the corresponding chemical composition, $C(x)$, for a given component, as e.g. for $^{57}$Co and Co in Fig. 6c. Then, the tracer diffusion coefficient, $D^*(\bar{C})$, as a function of the composition $\bar{C}$ is \cite{Murch1, Murch2},

\begin{equation}
 D^*(\bar{C}) = - \frac{ \frac{\bar{x}+a}{2t} - \frac{G(\bar{x})}{\bar{C}}}{\left. \frac{\partial \ln C^*}{\partial x} \right|_{x=\bar{x}} - \left. \frac{\partial \ln C}{\partial x}\right| _{x=\bar{x}}}\label{eq:BM-D}
\end{equation}

\noindent where the function $G$ is proportional to the flux of the given atoms and is determined as

\begin{equation}
 G(\bar{x}) = \frac{C^+-C^-}{2t} \left[ \left(1-\bar{Y} \right) \int_{-\infty}^{\bar{x}} Y dx +
                                        \bar{Y} \int^{+\infty}_{\bar{x}} \left(1-Y \right) dx \right]\label{eq:BM-G}
\end{equation}

\noindent Here $C^+$ and $C^-$ are the component concentrations at the right and left ends of the diffusion couple ($x = \pm \infty$), respectively, $a$ is a constant, which corresponds to the shift of the position of the Matano plane (given by the analysis of the chemical diffusion profile) with respect to the tracer concentration profile. The value of $a$ is determined by the condition that the Boltzmann analysis is applicable for both chemical and tracer diffusion profiles \cite{Murch1}, practically, by the fact that $D^*(C)>0$ for all concentrations $C$. The coordinate $\bar{x}$ corresponds to the location on the diffusion path where the chemical concentration of the given element is $\bar{C}$. $Y =(C-C^-)/(C^+-C^-)$ is the normalized concentration. Note that a constant molar volume is assumed (though this assumption is not strict and variable molar volume could be used \cite{Murch1}).

As a result, the concentration dependence of the tracer diffusion coefficient, $D^*(C)$, in a multicomponent alloy is determined without a general need to estimate the interdiffusion coefficient along the given path.

\begin{figure}[ht]
\footnotesize
 \begin{minipage}[c]{0.49\textwidth}
 a)\includegraphics[width=0.95\textwidth]{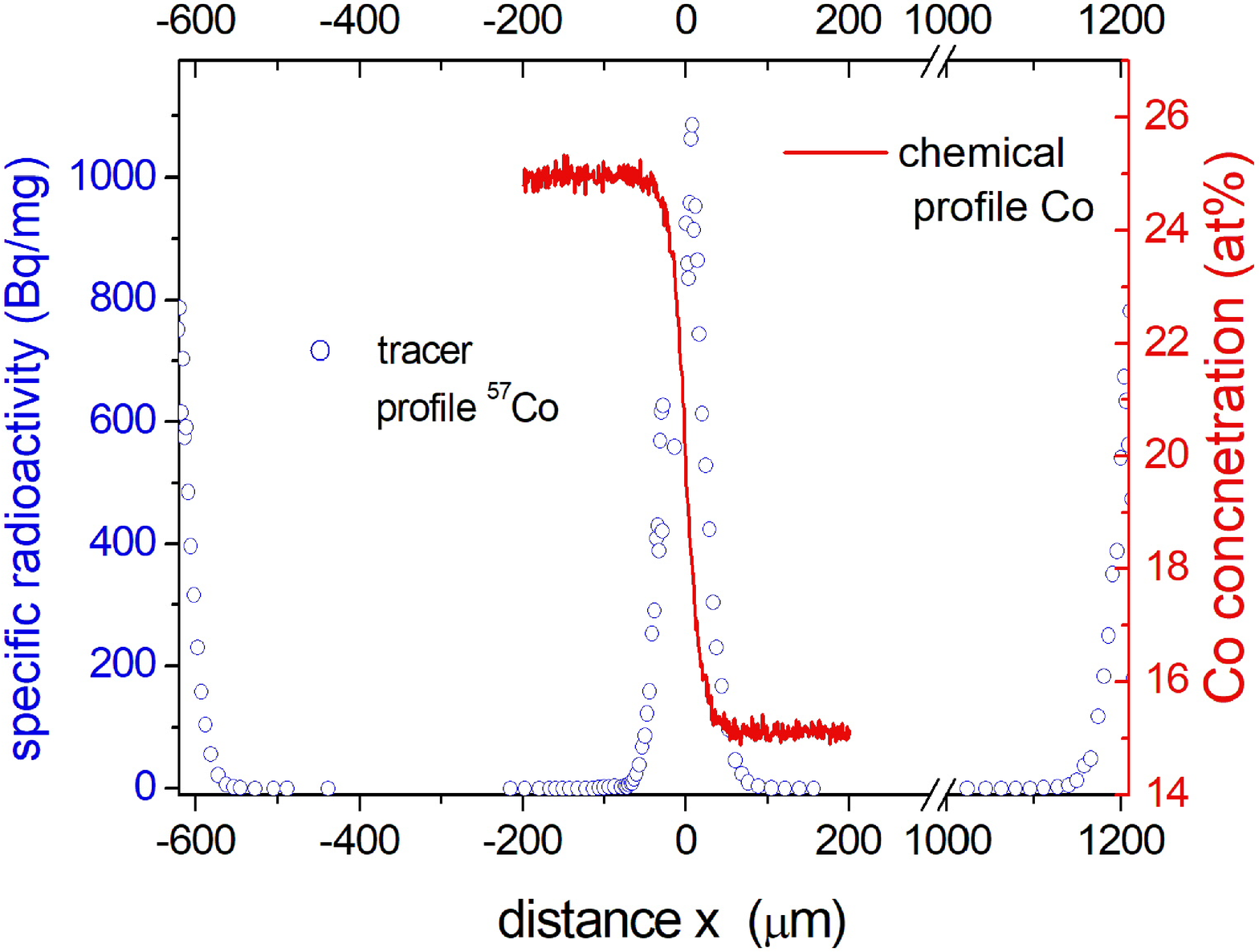}\hspace{0.3cm}
 \end{minipage}
 \hfill
 \begin{minipage}[c]{0.49\textwidth}
 b)\includegraphics[width=0.95\textwidth]{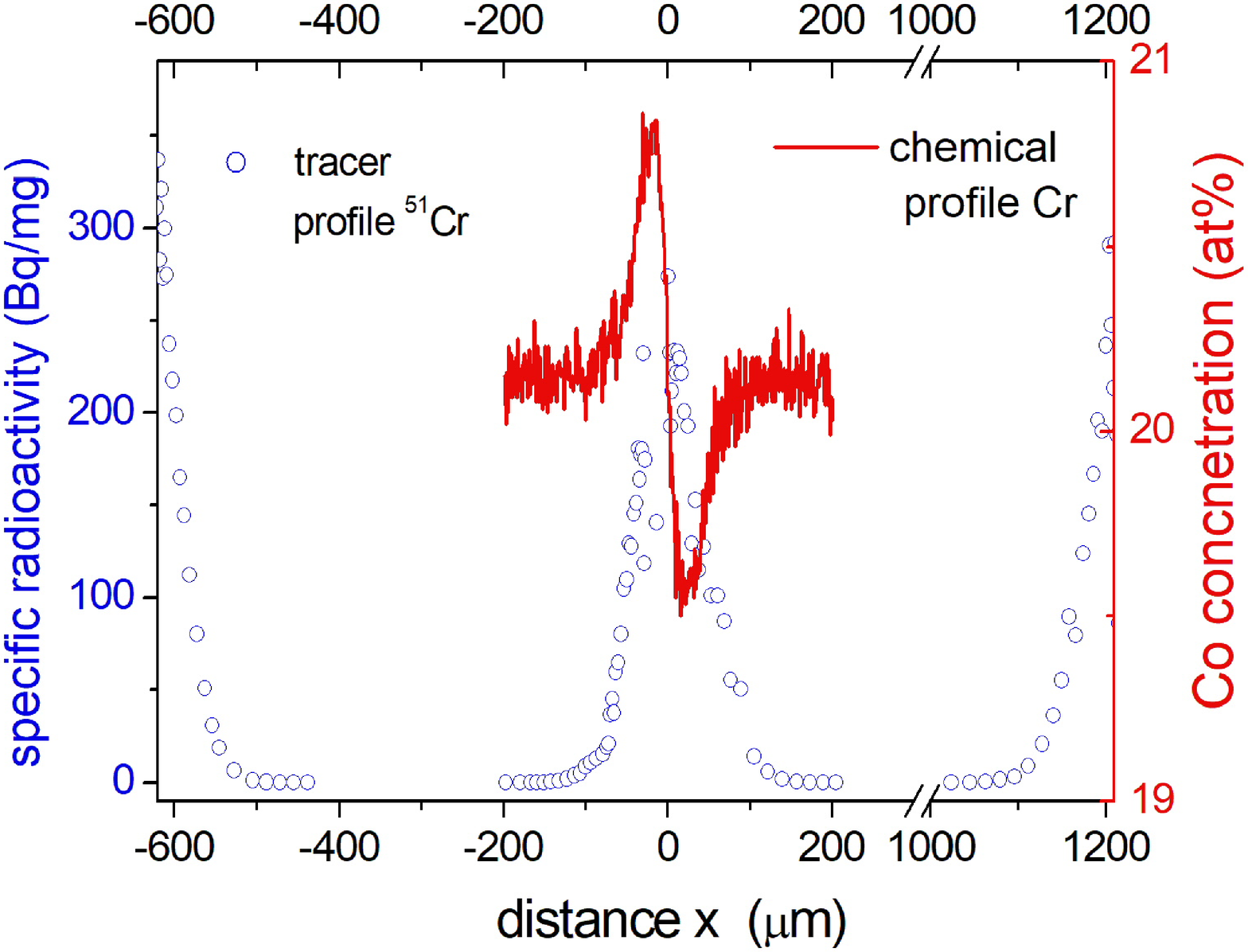}\hspace{0.3cm}
 \end{minipage}

 \begin{minipage}[c]{0.49\textwidth}
 c)\includegraphics[width=0.95\textwidth]{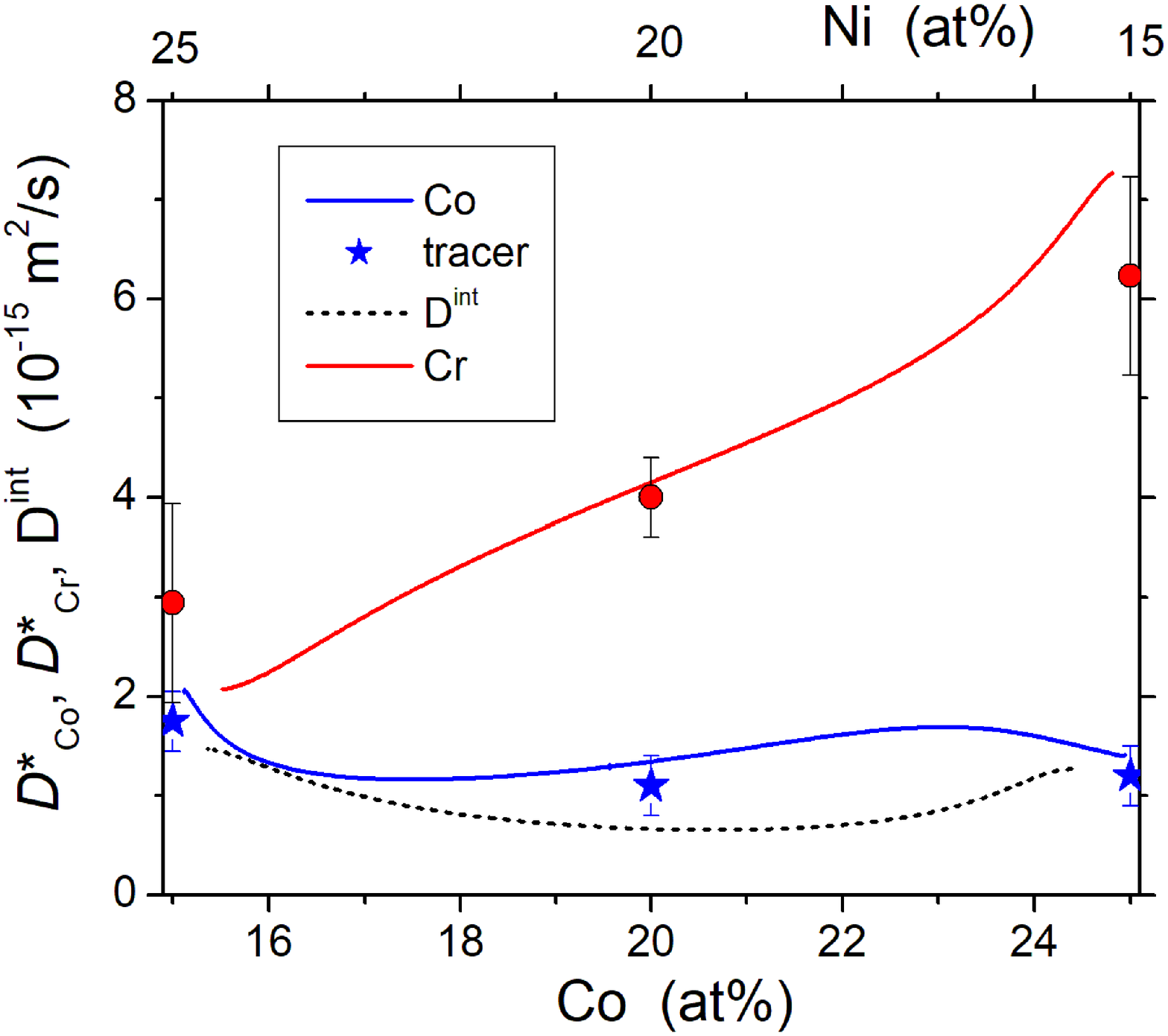}\hspace{0.3cm}
 \end{minipage}
 \hfill
 \begin{minipage}[c]{0.49\textwidth}
  \caption{\footnotesize Tracer (symbols) and chemical (lines) profiles measured for Co (a) and Cr (b) in the $^{51}$Cr+$^{57}$Co/ CrFeMnCo$_{15}$Ni$_{25}$/ $^{51}$Cr+$^{57}$Co/ CrFeMnCo$_{25}$Ni$_{15}$/$^{51}$Cr+$^{57}$Co tracer-interdiffusion couple \cite{Daniel-BM} and the determined tracer diffusion coefficients (c).}\label{fig:BMexp}
 \end{minipage}
\end{figure}

In Fig.~\ref{fig:BMexp} the preliminary data for the $^{51}$Cr+$^{57}$Co/ CrFeMnCo$_{15}$Ni$_{25}$/ $^{51}$Cr+$^{57}$Co/ CrFeMnCo$_{25}$Ni$_{15}$/$^{51}$Cr+$^{57}$Co tracer-interdiffusion couple measured at $T=1373$~K  \cite{Daniel-BM} are shown.

A good agreement of the directly measured tracer diffusion coefficients in the end-members (symbols) and those determined using Equations~(\ref{eq:BM-D}) and (\ref{eq:BM-G}) (solid lines) is seen.

Furthermore, the interdiffusion profile for Co (and correspondingly Ni) atoms can be processed as described above for the pseudo-binary system and the determined interdiffusion coefficient, $\widehat{D}^{int}(C)$, dashed line, is also plotted in Fig.\ref{fig:BMexp}c. Unfortunately the tracer diffusion coefficient $D^*_{\rm Ni}$ was not measured in Ref. \cite{Daniel-BM}, since $^{64}$Ni is a $\beta$-emitter and its decays cannot be measured by the $\gamma$-spectrometry and a further independent measurements is necessary, that in fact is a subject of planned research. 

However we may may check the validity of the Darken-Manning relation, Eq.~(\ref{eq:DM}), at the position of the Matano plane which correspond to the equiatomic composition Co$_{20}$Cr$_{20}$Fe$_{20}$Mn$_{20}$Ni$_{20}$ using the published tracer diffusion data in Ref. \cite{M-Ni}. A simple estimation yields that the product of the vacancy wind factor $\Omega$ and the thermodynamic factor $\Phi$ is $\Omega \cdot \Phi \approx 1.1 \pm 0.1$ \cite{Daniel-BM}. Thus we conclude that the CoCrFeMnNi HEA is almost ideal from the thermodynamic point of view at least at $T=1373$~K, as it is expected. Though, a strong up-hill diffusion for the components which are expected to develop no concentration profiles (in view of the same concentration on the both ends of the couple) was measured, see e.g. Fig.\ref{fig:BMexp}b for Cr. A similar behavior with a counterbalance of the up-hill diffusion was measured for Mn and practically flat profile was recorded for Fe \cite{Daniel-BM}. Thus, a strong coupling of kinetic terms is expected. The appearance of the cross-correlation terms is discussed in Ref.~\cite{Daniel-BM}. Here we note that Co and Ni reveal similar diffusion rates in CoCrFeMnNi at 1373~K -- the diffusion coefficients differ less than by a factor of two, see Fig.~\ref{fig:HEAs}b. Generally, one might not expect the appearance of the up-hill diffusion in such conditions, though the direct experiments substantiate the opposite behavior. We conclude that the given couple, being almost ideal from a thermodynamic point of view since $\Phi \approx 1$, is non-ideal and reveals strong cross-correlations between diffusion fluxes of different atoms. 

\begin{figure}[t]
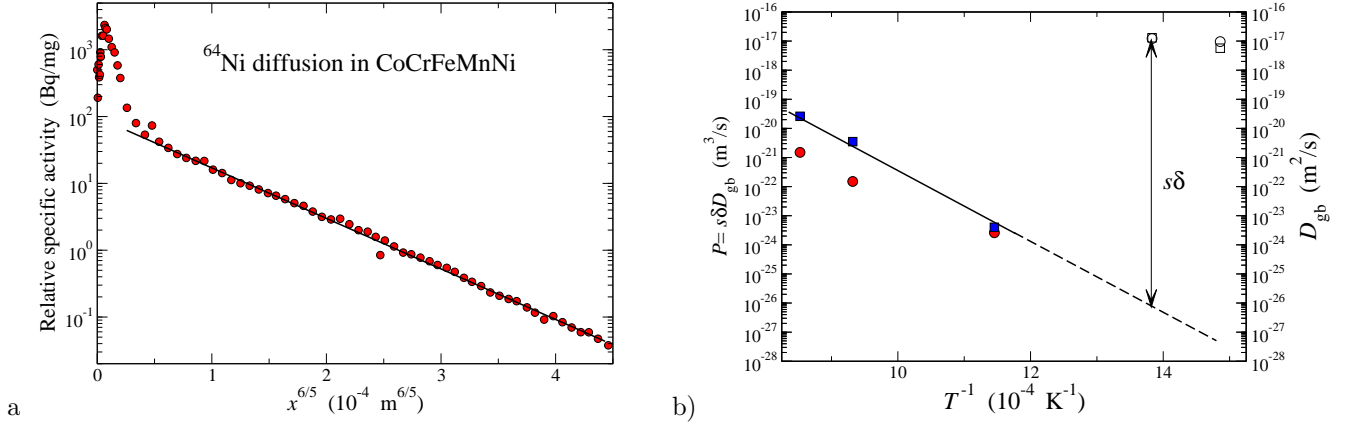

\centering
\footnotesize
\begin{minipage}[b]{0.48\textwidth}
a) \includegraphics[height=5.5cm]{figure13a}
\end{minipage}
\hfill
\begin{minipage}[b]{0.48\textwidth}
 b) \includegraphics[height=5.5cm]{figure13b}
\end{minipage}

\caption{\footnotesize An example of penetration profiles measured for Ni grain boundary diffusion in CoCrFeMnNi at 1173~K \cite{M-GB} (a) and the grain boundary diffusivities measured in CoCrFeNi (circles) and CoCrFeMnNi (squares) in the B-type (filled symbols) and C-type (open symbols) regimes. An estimate of the product $s\cdot \delta$ of grain boundary segregation factor $s$ and the grain boundary width $\delta$ is sketched.}\label{fig:GB}\normalsize
\end{figure}

\section{Grain boundary diffusion}
A direct advantage of the radiotracer technique is that not only bulk, but short-circuit diffusion, if it is present as in a polycrystalline material, can conveniently by measured, too, see Fig.~\ref{fig:HEAs}a. As a further example, in Fig.~\ref{fig:GB}a the penetration profile measured for Ni diffusion in CoCrFeMnNi at 1173~K \cite{M-GB} is shown. The penetration profile was measured at relatively high temperature under the so-called B-Type kinetic conditions after Harrison's classification \cite{Har}. In this case it is not the grain boundary di ffusion coefficient, $D_\gb$, which is determined by the corresponding slope of the grain boundary diffusion-related branch of the concentration profile, see the solid line in Fig.~\ref{fig:GB}a, but the so-called triple product $P$, $P = s \cdot \delta \cdot D_\gb$. Here $s$ is the grain boundary segregation factor and $\delta$ is the grain boundary width \cite{Paul2014}.

\begin{figure}[t]
\begin{center}
\includegraphics[width=0.8\textwidth]{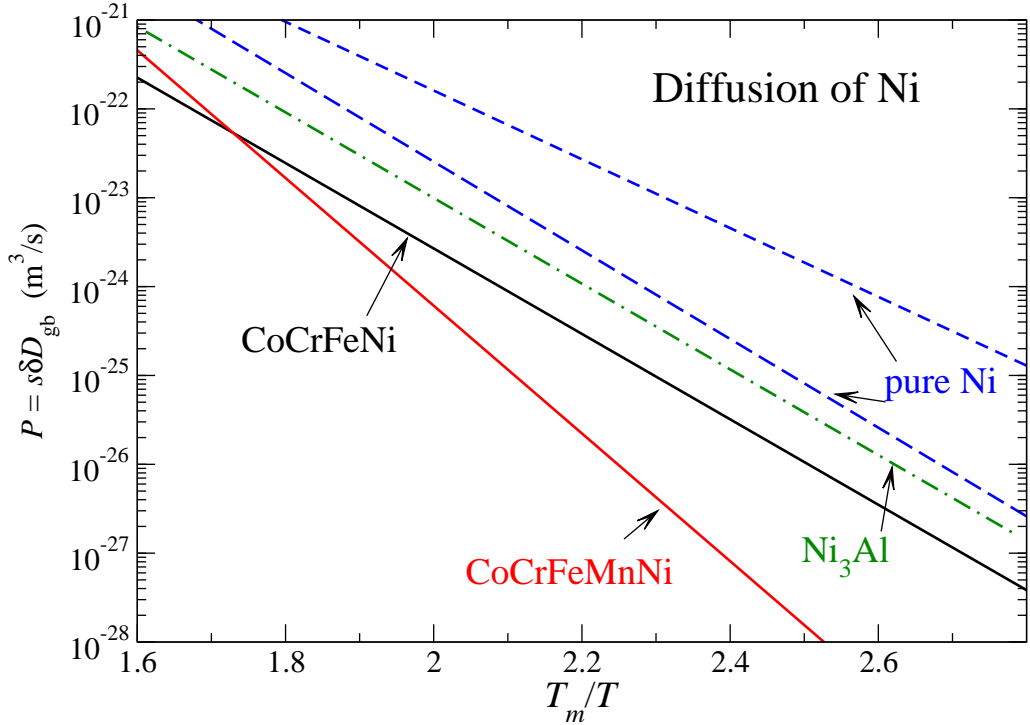}
\end{center}
\caption{\footnotesize The triple products $P$, $P = s\cdot \delta \cdot D_\gb$, of grain boundary diffusion of Ni in CoCrFeNi \cite{M-GB} (black solid line), CoCrFeMnNi \cite{M-GB} (red solid line), pure Ni \cite{Ni-Ger} (99.999~wt.\%, blue short-dashed line), less pure Ni \cite{Ni-Dar} (99.6~wt.\%, blue short-dashed line), and stoichiometric $L1_2$-ordered Ni$_3$Al alloy \cite{Ni3Al-GB} (green dot-dashed line) as function of the inverse homologous temperature scale, $T_m/T$. $T_m$ is the melting point of the corresponding compound.}\label{fig:GB-comp}\normalsize
\end{figure}

At lower temperatures, Ni grain boundary diffusion was measured in the so-called C-type regime after Harrison's classification and the Ni grain boundary diffusion coefficients, $D_\gb$, were determined \cite{M-GB}. In Fig.~\ref{fig:GB}b the measured data for both CoCrFeNi and CoCrFeMnNi HEAs are shown by plotting both triple products $P$, filled symbols and left ordinate, and the grain boundary diffusion coefficients, open symbols and right ordinate, in the same diagram. An extrapolation of the measured triple products to the lower temperatures substantiates that there is a gap between the $P$ and $D_\gb$ values which by definition is equal to the product $s\cdot \delta$, see Fig.~\ref{fig:GB}b, and one may estimate that $s\cdot \delta \approx 0.5$~nm. Since no segregation of Ni atoms to grain boundaries were found in Ref. \cite{M-GB}, we assume that $s=1$ and the diffusion data of Ref. \cite{M-GB} verify that the grain boundary width in FCC HEAs is about 0.5~nm. This value agrees perfectly with similar measurements on Ni diffusion in pure Ni and other FCC metals and alloys \cite{Ni-Dar, Ni-Ger}.

In Fig.~\ref{fig:GB-comp} the determined triple products for Ni diffusion in CoCrFeNi and CoCrFeMnNi HEAs are compared with the data on Ni grain boundary diffusion in Ni of different purity levels \cite{Ni-Dar, Ni-Ger} by using the homologous temperature scale $T_m/T$. One recognizes that Ni grain boundary diffusion reveals a certain retardation in HEAs with respect to that in pure Ni. However, this retardation is similar to that induced by residual impurities in Ni or by the $L1_2$ ordering in Ni$_3$Al, Fig.~\ref{fig:GB-comp}, and it depends strongly on the temperature.

Having measured both bulk and grain boundary self-diffusion rates one may use the Borisov semi-empirical relation \cite{Borisov} and estimate the grain boundary energy, $\gamma_\gb$, as

\begin{equation}
 \gamma_\gb = \frac{RT}{2a_0^2 N_A} \ln \left( \frac{D_\gb}{D_\vv} \right) 
\end{equation}

\begin{figure}[t]
\begin{minipage}[b]{0.7\textwidth}
\includegraphics[width=0.95\textwidth]{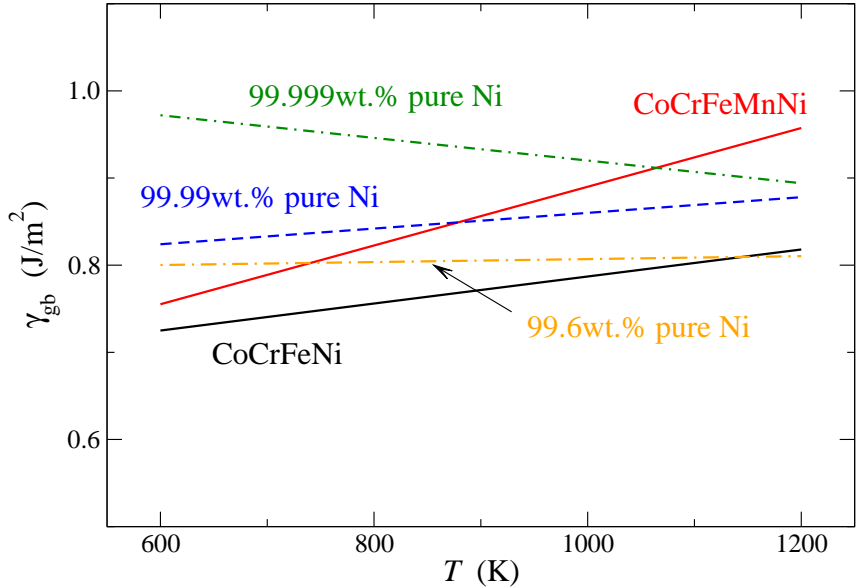}
\end{minipage}
\hfill
\begin{minipage}[b]{0.28\textwidth}
\caption{\footnotesize Grain boundary energies as determined via grain boundary diffusion measurements and the Borisov phenomenological approach in CoCrFeNi and CoCrFeMnNi \cite{M-GB} in comparison to the values determined for Ni of different purity levels \cite{Ni-Dar, Ni-Ger}.}\label{fig:GBenergy}\normalsize
\end{minipage}
\end{figure}

\noindent Here $a_0$ is the lattice parameter and $N_A$ is the Avogadro's number. Guiraldenq \cite{Guir} extended this concept to substitutional binary alloys and based on the reasoning in Ref. \cite{Ni-Dar}, Vaidya et al. \cite{M-GB} have proposed to apply this relation to HEAs, too. The determined grain boundary energies in CoCrFeNi and CoCrFeMnNi are plotted in Fig.~\ref{fig:GBenergy}, solid lines.

In Fig.~\ref{fig:GBenergy} the estimated values of grain boundary energy of HEAs are compared to those determined for Ni of different purity levels \cite{Ni-Dar, Ni-Ger}. The data suggest that the temperature dependence of grain boundary energy in CoCrFeMnNi is more pronounced as that in pure Ni. 

\section{Summary and Conclusions}

In the present review, a basic understanding of a complex diffusion behavior in HEAs is presented. Although being not 'sluggish' at all, the careful diffusion measurements discovered  a lot of exciting results and the term 'sluggish' has to be considered with a historical respect. In fact, this myth played a key role in initiating of the enormous interest to the field.

It is not a mysterious 'sluggish diffusion' which makes HEAs attractive for diffusional applications. HEAs offer an attractive playground for testing, verification and developing the new concepts of measurements and/or assessment of diffusion in multicomponent alloys, it is a perfect tool to go beyond linear approximations for database development and to validate the immense importance of kinetic cross-correlations between various elements.

Historically, the statement of 'sluggish diffusion' attracted increased attention of various diffusional groups worldwide (and allowed to receive fundings) at the initial period, but rigorous measurements have proven an elusiveness of this statement as a "core effect" of HEAs and provided a shift of paradigms to extended analysis of the complex diffusion phenomena in the multi-principal element alloys.

\section{Unresolved (open) problems}

To the author's opinion, there are several issues which have to be addressed in order to clarify the diffusion behavior in the high-entropy alloys.

\begin{itemize}
 \item To the date, diffusion in FCC crystalls have experimentally be measured and a moderate, if any, impact of an increased complexity of the chemical envinroment of diffusion atoms on its transport has been observed. It is worth to investigate the impact of an increased number of alloying components on diffusion in BCC and HCP lattices. \\
 An anisotropy of diffusion in HCP lattice is expected and a less dense packed crystalline structure of BCC alloys may provide paths for enhance diffusion of fast jumping species.

 \item An impact of partial (sublattice) order on diffusion in HEAs is a further aspect worth to be studied. Diffusion in ordered intermetallic compounds was intensively investigated in the past \cite{Div-ch10} and an intricate impact of Fe alloying on diffusion in the B2-ordered (Ni,Fe)$_{50}$Al$_{50}$ compounds was, e.g., observed \cite{NiFeAl, NiFeAl2}. 

 \item Diffusion along phase boundaries represents a special issue which is specific for the compositionally complex alloys with two-phase microstructures. An impact of composition and element segregation on the short-circuit transport may be expected, as it was observed, e.g., for Nb diffusion along the $\alpha_2$/$\gamma$ interfaces in Ti aluminides \cite{Ti-apb}.
 
 \item Solute diffusion in HEAs is almost compeletly uninvestigated subject and one may expect some deviations from general rules observed for pure metals.
 
 \item Dedicated experimental studies on interdiffusion should be conducted examining the interactions between different components and their role on the interdiffusion fluxes and the evolution of the composition profiles. 
\end{itemize}

\noindent {\bf Acknowledgments}. Although a personal view is presented, the author acknowledge numerous fruitful discussions with colleagues, especially with Prof. G. Wilde (University M\"unster, Germany), Prof. G. Murch and Prof. I. Belova (Australia). This work was partially supported by the German Science Foundation (DFG), project DI 1419/13-1. 

\renewcommand{\refname}{}

\end{document}